\documentclass[10pt]{IEEEtran}
\usepackage{cite,graphicx,psfrag,amsmath,amssymb,subfigure,url,supertabular,color}
\newtheorem{theorem}{Theorem}

\newtheorem{lemma}{Lemma}

\def\CampCost{L}
\def\definedas{\triangleq}
\def\order{O}
\def\s{\mbox{'s}}
\def\boldp{p}
\def\bigp{P}
\def\boldw{w}
\def\bigw{W}
\def\kval{k}
\def\kvals{k}
\def\len{n}
\def\biglen{N}
\def\E{{\mathbb E}}
\def\P{{\mathbb P}}
\def\R{{\mathbb R}}
\def\Rp{{{\mathbb R}_+}}

\def\X{{\mathcal X}}

\def\lg{{\log_2}}

\newcommand{\defn}[0]{\it}
\begin{document}
\bibliographystyle{IEEEtran} \title{Optimal Prefix Codes for Infinite
Alphabets with Nonlinear Costs}
\author{Michael~B.~Baer,~\IEEEmembership{Member,~IEEE}%
\thanks{This work was supported in part by the National Science
Foundation (NSF) under Grant CCR-9973134 and the Multidisciplinary
University Research Initiative (MURI) under Grant DAAD-19-99-1-0215.
Part of this work was performed while the author was at Stanford
University.  This material was presented in part at the IEEE
International Symposium on Information Theory, Seattle, Washington,
USA, July 2006 and at the IEEE International Symposium on Information Theory,
Nice, France, June 2007}%
\thanks{The author is with Ocarina Networks, Inc., 42 Airport Parkway, San Jose, CA  95110-1009  USA (e-mail:{\color{white}{i}}calbear{\color{black}{@}}{\bf \tiny \.{1}}eee.org).}
\thanks{This work has been submitted to the IEEE for possible publication. Copyright may be transferred without notice, after which this version may no longer be accessible.}}
\markboth{IEEE Transactions on Information Theory}{Optimal Prefix Codes for Infinite Alphabets with Nonlinear Costs}
\maketitle

\begin{abstract}
Let $\bigp = \{\boldp(i)\}$ be a measure of strictly positive
probabilities on the set of nonnegative integers.  Although the
countable number of inputs prevents usage of the Huffman algorithm,
there are nontrivial $\bigp$ for which known methods find a source
code that is optimal in the sense of minimizing expected codeword
length.  For some applications, however, a source code should instead
minimize one of a family of nonlinear objective functions,
$\beta$-exponential means, those of the form $\log_a \sum_i \boldp(i)
a^{\len(i)}$, where $\len(i)$ is the length of the $i$th codeword and
$a$ is a positive constant.  Applications of such minimizations
include a novel problem of maximizing the chance of message receipt in
single-shot communications ($a<1$) and a previously known problem of
minimizing the chance of buffer overflow in a queueing system ($a>1$).
This paper introduces methods for finding codes optimal for
such exponential means.  One method applies to geometric
distributions, while another applies to distributions with lighter
tails.  The latter algorithm is applied to Poisson distributions and
both are extended to alphabetic codes, as well as to minimizing
maximum pointwise redundancy.  The aforementioned application of
minimizing the chance of buffer overflow is also considered.
\end{abstract}


\IEEEpeerreviewmaketitle

\section{Introduction, Motivation, and Main Results} 
\label{intro} 

If probabilities are known, optimal lossless source coding of
individual symbols (and blocks of symbols) is usually done using David
Huffman's famous algorithm\cite{Huff}.  There are, however, cases that
this algorithm does not solve.  Problems with an
infinite number of possible inputs --- e.g., geometrically-distributed
variables --- are not covered.  Also, in some instances, the
optimality criterion --- or {\defn penalty} --- is not the linear
penalty of expected length.  Both variants of the problem have been
considered in the literature, but not simultaneously.  This paper 
discusses cases which are both infinite and nonlinear.

An infinite-alphabet source emits symbols drawn from the alphabet
$\X_\infty = \{0, 1, 2, \ldots \}$.  (More generally, we use $\X$ to
denote an input alphabet whether infinite or finite.)  Let $\bigp =
\{\boldp(i)\}$ be the sequence of probabilities for each symbol, so
that the probability of symbol $i$ is $\boldp(i) > 0$.  The source
symbols are coded into binary codewords.  The codeword $c(i) \in
\{0,1\}^*$ in code $C$, corresponding to input symbol~$i$, has length
$\len(i)$, thus defining length distribution~$\biglen$.  Such codes
are called {\defn integer codes} (as in, e.g., \cite{YaQi}).

Perhaps the most well-known integer codes are the codes derived by
Golomb for geometric distributions\cite{Golo,GaVV}, and many other
types of integer codes have been considered by others\cite{Abr01}.
There are many reasons for using such integer codes rather than codes
for finite alphabets, such as Huffman codes.  The most obvious use is
for cases with no upper bound --- or at least no known upper bound ---
on the number of possible items.  In addition, for many cases it is
far easier to come up with a general code for integers rather than a
Huffman code for a large but finite number of inputs.  Similarly, it
is often faster to encode and decode using such well-structured codes.
For these reasons, integer codes and variants of them are widely used
in image and video compression standards\cite{WSBL, WSS}, as well as
for compressing text, audio, and numerical data.

To date, the literature on integer codes has considered only finding
efficient uniquely decipherable codes with respect to minimizing
expected codeword length $\sum_i \boldp(i) \len(i)$.  Other utility
functions, however, have been considered for finite-alphabet codes.
Campbell~\cite{Camp} introduced a problem in which the penalty to
minimize, given some continuous (strictly) monotonic increasing {\defn
cost function} $\varphi(x):\Rp \rightarrow \Rp$, is 
$$
\CampCost(\bigp,\biglen,\varphi) = \varphi^{-1}\left(\sum_i \boldp(i)
\varphi(\len(i))\right) 
$$ and specifically considered the exponential subcases with exponent $a>1$:
\begin{equation} 
\CampCost_a(\bigp,\biglen) \definedas \log_a \sum_i \boldp(i) a^{\len(i)} 
\label{ExpCost} 
\end{equation} 
that is, $\varphi(x) = a^x$.  Note that minimizing penalty $\CampCost$
is also an interesting problem for $0<a<1$ and approaches the standard
penalty $\sum_i \boldp(i) \len(i)$ for $a \rightarrow 1$\cite{Camp}.
While $\varphi(x)$ decreases for $a<1$, one can map decreasing
$\varphi$ to a corresponding increasing function $\tilde{\varphi}(l) =
\varphi_{\max} - \varphi(l)$ (e.g., for $\varphi_{\max} = 1$) without
changing the penalty value.  Thus this problem, equivalent to
maximizing $\sum_i \boldp(i) a^{\len(i)}$, is a subset of those
considered by Campbell.  All penalties of the form (\ref{ExpCost}) are
called $\beta$-exponential means, where $\beta = \lg
a$\cite[p.~158]{AcDa}.

Campbell noted certain properties for $\beta$-exponential means, but
did not consider applications for these means.  Applications were
later found for the problem with $a>1$ \cite{Jeli,Humb2,BlMc};
these applications all relate to a buffer overflow problem
discussed in Section~\ref{application}.

Here we introduce a novel application for problems of the form $a<1$.
Consider a situation related by Alfred R\'{e}nyi, an ancient scenario
in which a rebel fortress was besieged by Romans.  The rebels' only
hope was the knowledge gathered by a mute, illiterate spy, one who
could only nod and shake his head \cite[pp.~13-14]{Reny}.  This
apocryphal tale --- based upon a historical siege --- is the premise
behind the Hungarian version of the spoken parlor game Twenty
Questions.  A modern parallel in the 21\textsuperscript{st} century
occurred when Russian forces gained the knowledge needed to defeat
hostage-takers by asking hostages ``yes'' or ``no'' questions over
mobile phones\cite{MSN,Tar}.

R\'{e}nyi presented this problem in narrative form in order to
motivate the relation between Shannon entropy and binary prefix
coding.  Note however that Twenty Questions, traditional prefix
coding, and the siege scenario actually have three different
objectives.  In Twenty Questions, the goal is to be able to determine
the symbol (i.e., the item or message) by asking at most twenty
questions.  In prefix coding, the goal is to minimize the expected
number of questions --- or, equivalently, bits --- necessary to
determine the message.  For the siege scenario, the goal is survival;
that is, assuming partial information is not useful, the besieged
would wish to maximize the probability that the message is
successfully transmitted within a certain window of opportunity.  When
this window closes --- e.g., when the fortress falls --- the
information becomes worthless.  An analogous situation occurs when a
wireless device is losing power or is temporarily within range of a
base station; one can safely assume that the channel, when available,
will transmit at the lowest (constant) bitrate, and will be lost after a
nondeterministic time period.

Assume that the duration of the window of opportunity is independent
of the communicated message and is memoryless, the latter being a
common assumption --- due to both its accuracy and expedience --- of
such stochastic phenomena.  Memorylessness implies that the window
duration is distributed exponentially.  Therefore, quantizing time in
terms of the number of bits $T$ that we can send within our window,
$$\P(T = t) = (1-a)a^t, ~ t = 0, 1, 2, \ldots $$ with known positive parameter
$a<1$.  We then wish to maximize the probability of success, i.e., the
probability that the message length does not exceed the quantized
window length:
\begin{eqnarray*}
\P[\len(X) \leq T] &=& \sum_{t=0}^\infty \P(T=t) \cdot \P[\len(X) \leq t] \\
&=& \sum_{t=0}^\infty (1-a)a^t \cdot \sum_{i \in \X} p(i) 1_{\len(i) \leq t} \\
&=& \sum_{i \in \X} p(i) \cdot (1-a) \sum_{t=\len(i)}^\infty a^t \\
&=& \sum_{i \in \X} p(i) a^{\len(i)} \cdot (1-a) \sum_{t=0}^\infty a^t \\
&=& \sum_{i \in \X} p(i) a^{\len(i)}
\end{eqnarray*}
where $1_{\len(i) \leq t}$ is $1$ if $\len(i) \leq t$, $0$ otherwise.
Minimizing (\ref{ExpCost}) is an equivalent objective.

Note that this problem can be constrained or otherwise modified for
the application in question.  For example, in some cases, we might
need some extra time to send the first bit, or, alternatively, the
window of opportunity might be of at least a certain duration,
increasing or reducing the probability that no bits can be sent,
respectively.  Thus we might have
$$ \P(T = t) = \left\{
\begin{array}{ll}
t_0,& t = 0 \\
(1-t_0)(1-a)a^{t-1},& t = 1, 2, \ldots 
\end{array}
\right.
$$ for some~$t_0 \in (0,1)$.  In this case, 
$$\P[\len(X) \leq T] = \frac{(1-t_0)}{a} \sum_{i \in \X} p(i)
a^{\len(i)}$$ and the maximizing code is identical to that of the more
straightforward case.  Likewise, if we need to send multiple messages,
the same code maximizes the expected number of independent messages we can
send within the window, due to the memoryless property.

We must be careful regarding the meaning of an ``optimal code'' when
there are an infinite number of possible codes under consideration.
One might ask whether there must exist an optimal code or if there can
be an infinite sequence of codes of decreasing penalty without any
code achieving the limit penalty value.  Fortunately the answer is the
former, the proof being a special case of Theorem~2 in~\cite{Baer06}
(a generalization of the result for the expected-length
penalty\cite{LTZ}).  The question is then how to find one of these
optimal source codes given parameter $a$ and probability
measure~$\bigp$.

As in the linear case, a general solution for (\ref{ExpCost}) is not
known for general $\bigp$ over a countably infinite number of events,
but methods and properties for finite numbers of events --- discussed
in the next section --- can be used to find optimal codes for certain
common infinite-item distributions.  In Section~\ref{geometric}, we
consider geometric distributions and find that Golomb codes are
optimal, although the optimal Golomb code for a given probability mass
function varies according to $a$.  The main result of
Section~\ref{geometric} is that, for $\boldp_\theta(i) =
(1-\theta)\theta^i$ and $a \in \Rp$, G$\kval$, the Golomb code with
parameter $\kval$, is optimal, where $$\kval = \max\left(1,
\left\lceil -\log_\theta a -\log_\theta (1+\theta)
\right\rceil\right).$$ In Section~\ref{other}, we consider
distributions that are relatively light-tailed, that is, that decline
faster than certain geometric distributions.  If there is a
nonnegative integer $r$ such that for all $j>r$ and $i<j$,
$$\boldp(i) \geq \max\left(\boldp(j), \sum_{k=j+1}^\infty \boldp(k)
a^{k-j}\right)$$ then an optimal binary prefix code tree exists which
consists of a unary code tree appended to a leaf of a finite code
tree.  A specific case of this is the Poisson distribution,
$\boldp_\lambda(i)=\lambda^i e^{-\lambda}/i!$, where $e$ is the base
of the natural logarithm ($e \approx 2.71828$).  We show that in this
case the aforementioned $r$ is given by $r = \max(\lceil 2 a \lambda
\rceil - 2, \lceil e \lambda \rceil - 1)$.  An application, that of
minimizing probability of buffer overflow, as in~\cite{Humb2}, is
considered in Section~\ref{application}, where we show that the
algorithm developed in \cite{Humb2} readily extends to coding
geometric and light-tailed distributions.  Section~\ref{nonexp}
discusses the maximum pointwise redundancy penalty, which has a
similar solution for light-tailed distributions and for which the
Golomb code G$\kval$ with $\kval = \lceil -1/\lg \theta \rceil$ is
optimal for with geometric distributions.  We conclude with some
remarks on possible extensions to this work.

Throughout the following, a set or sequence of items $x(i)$ is
represented by its uppercase counterpart, $X$.  A glossary of terms is
given in Appendix~\ref{glossary}.

\section{Background: Finite Alphabets}
\label{background} 
 
If a finite number of events comprise $\bigp$ (i.e., $|\X|<\infty$),
the exponential penalty (\ref{ExpCost}) is minimized using an
algorithm found independently by Hu {\it et al.}~\cite[p.~254]{HKT},
Parker \cite[p.~485]{Park}, and Humblet
\cite[p.~25]{Humb0},\cite[p.~231]{Humb2}, although only the last of
these considered $a < 1$.  (The simultaneity of these lines of
research was likely due to the appearance of the first paper on
adapting the Huffman algorithm to a nonlinear penalty, $\max_i
(\boldp(i) + \len(i))$ for given $\boldp(i) \in \Rp$, in
1976\cite{Golu}.)  We will use this finite-alphabet
exponential-penalty algorithm in the sections that follow in order to
prove optimally for infinite distributions, so let us reproduce the
algorithm here:

\textbf{Procedure for Exponential Huffman Coding (finite alphabets):} 
This procedure finds the optimal code
whether $a>1$ (a minimization of the average of a growing exponential)
or $a<1$ (a maximization of the average of a decaying exponential).
Note that it minimizes (\ref{ExpCost}), even if the ``probabilities''
do not add to $1$.  We refer to such arbitrary positive inputs as
{\defn weights}, denoted by $\boldw(i)$ instead of~$\boldp(i)$:

\begin{enumerate}
\item Each item $i$ has weight $\boldw(i) \in \bigw_{\X}$, where $\X$
is the (finite) alphabet and $\bigw_{\X} = \{w(i)\}$ is the set of all
such weights.  Assume each item $i$ has codeword $c(i)$, to be
determined later.
\item Combine the items with the two smallest weights $\boldw(j)$ and
  $\boldw(k)$ into one compound item with the combined weight
  $\tilde{\boldw}(j) = a \cdot (\boldw(j) + \boldw(k))$.  This item
  has codeword $\tilde{c}(j)$, to be determined later, while item $j$ is
  assigned codeword $c(j) = \tilde{c}(j)0$ and $k$ codeword $c(k) =
  \tilde{c}(j)1$.  Since these have been assigned in terms of
  $\tilde{c}(j)$, replace $\boldw(j)$ and $\boldw(k)$ with
  $\tilde{\boldw}(j)$ in $\bigw_\X$ to form $\bigw_{\tilde{\X}}$.
\item Repeat procedure, now with the remaining codewords (reduced in
  number by $1$) and corresponding weights, until only one
  item is left.  The weight of this item is $\sum_i \boldw(i)
  a^{\len(i)}$.  All codewords are now defined by assigning the null
  string to this trivial item.
\end{enumerate}
This algorithm assigns a weight to each node of
the resulting implied code tree by having each item represented by a
node with its parent representing the items combined into its subtree,
as in Fig.~\ref{buildgolo}: If a node is a leaf, its weight is given
by the associated probability; otherwise its weight is defined
recursively as $a$ times the sum of its children.  This concept is
useful in visualizing both the coding procedure and its output.

Van Leeuwen implemented the Huffman algorithm in linear time (to input
size) given sorted weights in \cite{Leeu}, and this implementation was
extended to the exponential problem in \cite{Baer05} as follows:

\textbf{Two-Queue Implementation of Exponential Huffman Coding:}
The two-queue method of implementing the Huffman algorithm puts
nodes/items in two queues, the first of which is initialized with the
input items (eventual leaf nodes) arranged from head to tail in order
of nondecreasing weight, and the second of which is initially empty.
At any given step, a node with lowest weight among all nodes in both
queues is at the head of one of the two queues, and thus two
lowest-weighted nodes can be combined in constant time.  This compound
node is then inserted into (the tail of) the second queue, and the
algorithm progresses until only one node is left.  This node is the
root of the coding tree and is obtained in linear time.

The presentation of the algorithm in \cite{Baer05} did not include a
formal proof, so we find it useful to present one here:

\begin{lemma}
The two-queue method using the exponential combining rule
results in an optimal exponential Huffman code given a finite number
of input items.  
\label{twoqueue}
\end{lemma}

\begin{proof}
The method is clearly a valid implementation of the exponential
Huffman algorithm so long as both queues' sets of nodes remain in
nondecreasing order.  This is clearly satisfied prior to the first
combination step.  Here we show that, if nodes are in order at all
points prior to a given combination step, they must be in order at the
end of that step as well, inductively proving the correctness of the
algorithm.  It is obvious that order is preserved in the single-item
queue, since nodes are only removed from it, not added to it.  In the
compound-node queue, order is only a concern if there is already at
least one node in it at the beginning of this step, a step that
combines nodes we call node $i_{-1}$ and node $i_{-2}$.  If so, the
item at the tail of the compound-node queue at the beginning of the
step was two separate items, $i_{-3}$ and $i_{-4}$, at the beginning
of the prior step.  At the beginning of this prior step, all four
items must have been distinct --- i.e., corresponding to distinct sets
of (possibly combined) leaf nodes --- and, because the algorithm
chooses the smallest two nodes to combine, neither $i_{-3}$ nor
$i_{-4}$ can have a greater weight than either $i_{-1}$ or $i_{-2}$.
Thus --- since $a\cdot(\boldw(i_{-3})+\boldw(i_{-4})) \leq
a\cdot(\boldw(i_{-1})+\boldw(i_{-2}))$ and the node with weight
$a\cdot(\boldw(i_{-3})+\boldw(i_{-4}))$ is the compound node with the
largest weight in the compound-node queue at the beginning of the step
in question --- the queues remain properly ordered at the end of the
step in question.
\end{proof}

If $a < 0.5$, the compound-node queue will never have more than one
item.  At each step after the first, the sole compound item will be
removed from its queue since it has a weight less than the maximum
weight of each of the two nodes combined to create it, which in turn
is no greater than the weight of any node in the single-item queue.
It is replaced by the new (sole) compound item.  This extends to $a =
0.5$ if we prefer to merge combined nodes over single items of the
same weight.  Thus, any finite input distribution can be optimally
coded for $a \leq 0.5$ using a {\defn truncated unary code}, a
truncated version of the {\defn unary code}, the latter of which has
codewords of the form $\{1^j0 : j \geq 0\}$.  The truncated unary code
has identical codewords as the unary code except for the longest
codeword, which is of the form $\{1^{|\X|-1}\}$.  This results from
each compound node being formed using at least one single item (leaf).
Taking limits, informally speaking, results in a unary limit code.
Formally, this is a straightforward corollary of Theorem~\ref{tailthm}
in Section~\ref{other}.

If $a>0.5$, a code with finite penalty exists if and only if R\'{e}nyi
entropy of order $\alpha(a) = {(1+\lg a)}^{-1}$ is finite, as shown in
\cite{Baer06}.  It was Campbell who first noted the connection between
the optimal code's penalty, $\CampCost_a(\bigp,\biglen^*)$, and
R\'{e}nyi entropy
\begin{eqnarray*}
H_{\alpha}(\bigp) &\definedas& \frac{1}{1-\alpha} \lg \sum_{i \in \X} 
\boldp(i)^\alpha \\
\Rightarrow H_{\alpha(a)}(\bigp) &=& \frac{1+\lg a}{\lg a} 
\lg \sum_{i \in \X} \boldp(i)^{(1+\lg a)^{-1}} .
\end{eqnarray*}
This relationship is
$$H_{\alpha(a)}(\bigp) \leq \CampCost_a(\bigp,\biglen^*) <
H_{\alpha(a)}(\bigp) + 1$$ which should not be surprising given the
similar relationship between Huffman-optimal codes and Shannon
entropy\cite{Shan}, which corresponds to $a \rightarrow 1$ ($\alpha
\rightarrow 1$)\cite{Ren2,Camp}; due to this correspondence, Shannon
entropy is sometimes expressed as $H_1(\bigp)$.

\section{Geometric Distribution with Exponential Penalty}
\label{geometric}

Consider the geometric distribution $$\boldp_\theta(i) =
(1-\theta)\theta^i$$ for parameter $\theta \in (0,1)$.  This
distribution arises in run-length coding among other
circumstances\cite{Golo,GaVV}.

For the traditional linear penalty, a Golomb code with
parameter~$\kval$ --- or G$\kval$ --- is optimal for $\theta^\kval +
\theta^{\kval+1} \leq 1 < \theta^{\kval-1} + \theta^\kval$.  Such a
code consists of a unary prefix followed by a binary suffix, the
latter taking one of $\kval$ possible values.  If $\kval$ is a power
of two, all binary suffix possibilities have the same length;
otherwise, their lengths $\sigma(i)$ differ by at most $1$ and $\sum_i
2^{-\sigma(i)}=1$.  Binary codes such as these suffix codes are called
{\defn complete} codes.  This defines the Golomb code; for example,
the Golomb code for $\kval = 3$ is:
\begin{center}
$$
\begin{array}{rll}
\hline
\hline
i&\boldp(i)&c(i) \\
\hline
0&1-\theta&0~0 \\
1&(1-\theta)\theta&0~10 \\
2&(1-\theta)\theta^2&0~11 \\
3&(1-\theta)\theta^3&10~0 \\
4&(1-\theta)\theta^4&10~10 \\
5&(1-\theta)\theta^5&10~11 \\
6&(1-\theta)\theta^6&110~0 \\
7&(1-\theta)\theta^7&110~10 \\
8&(1-\theta)\theta^8&110~11 \\
9&(1-\theta)\theta^9&1110~0 \\
\vdots&\qquad \vdots&\qquad \vdots \\
\hline
\end{array}
$$
\end{center}
where the space in the code separates the unary prefix from the complete
suffix.  In general, codeword $j$ for G$\kval$ is of the form
$\{1^{\lfloor j/\kval \rfloor} 0 b(j \bmod \kval,\kval) : j \geq 0\}$,
where $b(j \bmod \kval, \kval)$ is a complete binary code for the $(j -
\kval \lfloor j/\kval \rfloor+1)$th of $\kval$ items.

It turns out that such codes are optimal for the exponential penalty:
\begin{theorem}
For $a \in \Rp$, if
\begin{equation}
\theta^\kval + \theta^{\kval+1} \leq \frac{1}{a} < \theta^{\kval-1} +
\theta^\kval
\label{ineq}
\end{equation}
for $\kval \geq 1$, then the Golomb code G$\kval$ is the optimal code
for $\bigp_\theta$.  If no such $\kval$ exists, the unary code G$1$ is
optimal.
\label{optgeo}
\end{theorem}

\textit{Remark:} This rule for finding an optimal Golomb G$\kval$ code
is equivalent to
$$\kval = \max\left(1, \left\lceil -\log_\theta a -\log_\theta
(1+\theta) \right\rceil\right).$$ This is a generalization of the
traditional linear result, which corresponds to $a \rightarrow 1$.
Cases in which the left inequality is an equality have multiple
solutions, as with linear coding; see, e.g., \cite[p.~289]{Goli2}.
The proof of the optimality of the Golomb code for exponential
penalties is somewhat similar to that of \cite{GaVV}, although it must
be significantly modified due to the nonlinearity involved.

Before proving Theorem~\ref{optgeo}, we need the following lemma:

\begin{lemma}
Consider a Huffman combining procedure, such as the exponential
Huffman coding procedure, implemented using the two-queue method presented in the previous section just prior to Lemma~\ref{twoqueue}.  Now consider a step at which the first (single-item)
queue is empty, so that remaining are only compound items, that is,
items representing internal nodes rather than leaves in the final
Huffman coding tree.  Then, in this final tree, the nodes corresponding to these compound items will be on
levels differing by at most one; that is, the nodes will form a
complete tree.  Furthermore, if $n$ is the number of items remaining
at this point, all items that finish at level $\lceil \lg n \rceil$
appear closer to the head of the (second, nonempty) queue than any
item at level $\lceil \lg n \rceil - 1$ (if any).
\label{thelemma}
\end{lemma}

\begin{proof}[Lemma~\ref{thelemma}]
We use an inductive proof, in which the base cases of one and two
compound items (i.e., internal nodes) are trivial.  Suppose the lemma is
true for every case with $n-1$ items for $n>2$, that is, that all
nodes are at levels $\lfloor \lg (n-1) \rfloor$ or $\lceil \lg (n-1)
\rceil$, with the latter items closer to the head of the queue than
the former.  Consider now a case with $n$ nodes.  The first step of
coding is to merge two nodes, resulting in a combined item that is
placed at the end of the combined-item queue.  Because it is at the
end of the queue in the reduced problem of size $n-1$, this combined node is at level
$\lfloor \lg (n-1) \rfloor$ in the final tree, and its children are at
level $1+\lfloor \lg (n-1) \rfloor = \lceil \lg n \rceil$.  If $n$ is
a power of two, the remaining items end up on level $\lg n = \lceil \lg
(n-1) \rceil$, satisfying this lemma.  If $n-1$ is a
power of two, they end up on level $\lg (n-1) = \lfloor \lg n \rfloor$,
also satisfying the lemma.  Otherwise, there is at least one item ending up at
level $\lceil \lg n \rceil = \lceil \lg (n-1) \rceil$ near the head of
the queue, followed by the remaining items, which end up at level
$\lfloor \lg n \rfloor = \lfloor \lg (n-1) \rfloor$.  In any case, the
lemma is satisfied for $n$ items, and thus, inductively, for any number of items.
\end{proof}

This lemma applies to any problem in which a two-queue Huffman algorithm provides an optimal solution, including the original Huffman
problem and the tree-height problem of \cite{Park}.  Here we apply the lemma to the exponential Huffman algorithm to prove Theorem~\ref{optgeo}:

\begin{figure*}
\psfrag{  0}{\mbox{\tiny $w(0)$}}
\psfrag{  1}{\mbox{\tiny $w(1)$}}
\psfrag{  2}{\mbox{\tiny $w(2)$}}
\psfrag{  3}{\mbox{\tiny $w(3)$}}
\psfrag{  4}{\mbox{\tiny $w(4)$}}
\psfrag{  5}{\mbox{\tiny $w(5)$}}
\psfrag{  6}{\mbox{\tiny $w(6)$}}
\psfrag{  7}{\mbox{\tiny $w(7)$}}
\psfrag{  8}{\mbox{\tiny $w(8)$}}
\psfrag{  9}{\mbox{\tiny $w(9)$}}
\psfrag{  10}{\mbox{\tiny $w(10)$}}
\psfrag{  11}{\mbox{\tiny $w(11)$}}
\psfrag{  12}{\mbox{\tiny $w(12)$}}
\psfrag{  13}{\mbox{\tiny $w(13)$}}
\psfrag{  14}{\mbox{\tiny $w(14)$}}
\psfrag{  15}{\mbox{\tiny $w(15)$}}
\psfrag{  16}{\mbox{\tiny $w(16)$}}
\psfrag{  17}{\mbox{\tiny $w(17)$}}
\psfrag{  18}{\mbox{\tiny $w(18)$}}
\psfrag{  19}{\mbox{\tiny $w(19)$}}
\psfrag{  20}{\mbox{\tiny $w(20)$}}
\psfrag{  21}{\mbox{\tiny $w(21)$}}
\psfrag{  22}{\mbox{\tiny $w(22)$}}
\begin{center}
\resizebox{14cm}{!}{\includegraphics{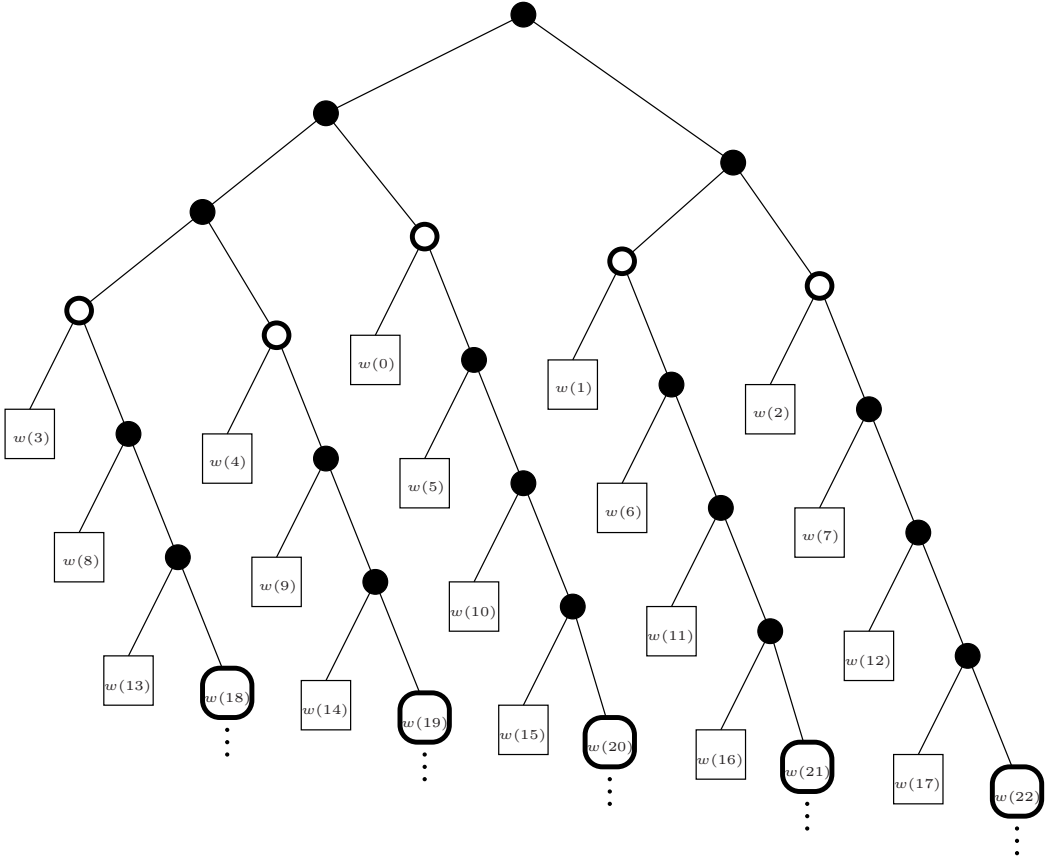}}
\caption{Formation of a Golomb code using a code for an $m$-reduced
source.  In this illustration, $m=17$ and $\kval=5$, and smaller weights are pictorially lower.  Weights are merged bottom-up, in a manner consistent with the exponential Huffman algorithm, first in separate (truncated) unary subtrees, then in a (five-leaf) complete tree.}
\label{buildgolo}
\end{center}
\end{figure*}

\begin{proof}[Theorem~\ref{optgeo}]
We start with an optimal exponential Huffman code for a sequence of
similar finite weight distributions.  These finite weight
distributions, called {\defn $m$-reduced geometric sources} $\bigw_m$,
are defined as:
$$
\boldw_m(i) \definedas \left\{
\begin{array}{ll}
(1-\theta)\theta^i,& 0 \leq i \leq m \\
\displaystyle
\frac{(1-\theta)a\theta^i}{1-a\theta^\kval},& m < i \leq m + \kval .\\
\end{array}
\right.
$$ where $\kval$ is as given in the statement of the theorem, or $1$
if no such $\kval$ exists.  

Weights $\boldw_m(0)$ through $\boldw_m(m)$ are decreasing, as are
$\boldw_m(m+1)$ through $\boldw_m(m+\kval)$.  Thus we can combine the
nodes with weights $\boldw_{m}(m)$ and $\boldw_m(m+\kval)$ if
$$\frac{(1-\theta)a\theta^{m+\kval}}{1-a\theta^\kval} \leq 
(1-\theta)\theta^{m-1}$$
and
$$\frac{(1-\theta)a\theta^{m+\kval-1}}{1-a\theta^\kval} >
(1-\theta)\theta^m \mbox{ or } \kval=1.$$ These conditions
are equivalent to the left and right sides, respectively, of
(\ref{ineq}).  Thus the combined item is
$$\boldw_{m-1}(m) = \frac{(1-\theta)a\theta^m}{1-a\theta^\kval}$$ 
and the code is reduced to the $\bigw_{m-1}$ case.

After merging the two smallest weights for $m=0$, the reduced source
is $$\boldw_{-1}(i) = \frac{(1-\theta)a\theta^i}{1-a\theta^\kval}, ~ 0
\leq i \leq \kval-1 .$$ For $\kval=1$ (including all instances of the
degenerate $a \leq 0.5$ case and all instances in which (\ref{ineq})
cannot be satisfied), this proves that the optimal tree is the
truncated unary tree.  Considering now only $\kval>1$ for $m \geq
\kval-1$, the two-queue algorithm assures that, when the problem is
reduced to weights $\{\boldw_{-1}(i)\}$, all corresponding nodes are
in the combined-item queue.  Lemma~\ref{thelemma} thus proves that
these nodes form a complete code.  The overall optimal tree for any
$m$-reduced code with $m \geq \kval-1$ is then a truncated Golomb
tree, as pictorially represented in Fig.~\ref{buildgolo}, where $m=17$
and $\kval=5$.  Note that $m+1$ is the number of leaves in common with
what we call the ``Golomb tree,'' the tree we show to be optimal for
the original geometric source.  The number of remaining leaves in the
truncated tree is~$\kval$, which is thus the number of distinct unary
subtrees in the Golomb tree.

Fig.~\ref{buildgolo} represents both the truncated and full Golomb
trees, along with how to merge the weights.  Squares represent items
to code, while circles represent other nodes of the tree.  Smaller
weights are below larger ones, so that items are merged as pictured.
Rounded squares are items $m+1$ through $m+\kval$, the items which are
replaced in the Golomb tree by unary subtrees, that is, subtrees
representing the unary code.  Other squares are items $0$ through $m$,
those corresponding to single items in the integer code.  White
circles are the leaves used for the complete tree.

\begin{figure*}[ht]
\psfrag{L-Ha}{$\bar{R}_a(\biglen_{\theta,a}^*,\bigp_\theta)$}
\psfrag{a}{$a$}
\psfrag{ag}{\mbox {\huge $a$}}
\psfrag{Theta}{$\theta$}
     \centering
     \subfigure[$a>1$]
	       { \label{apos} \includegraphics[width=.45\textwidth]{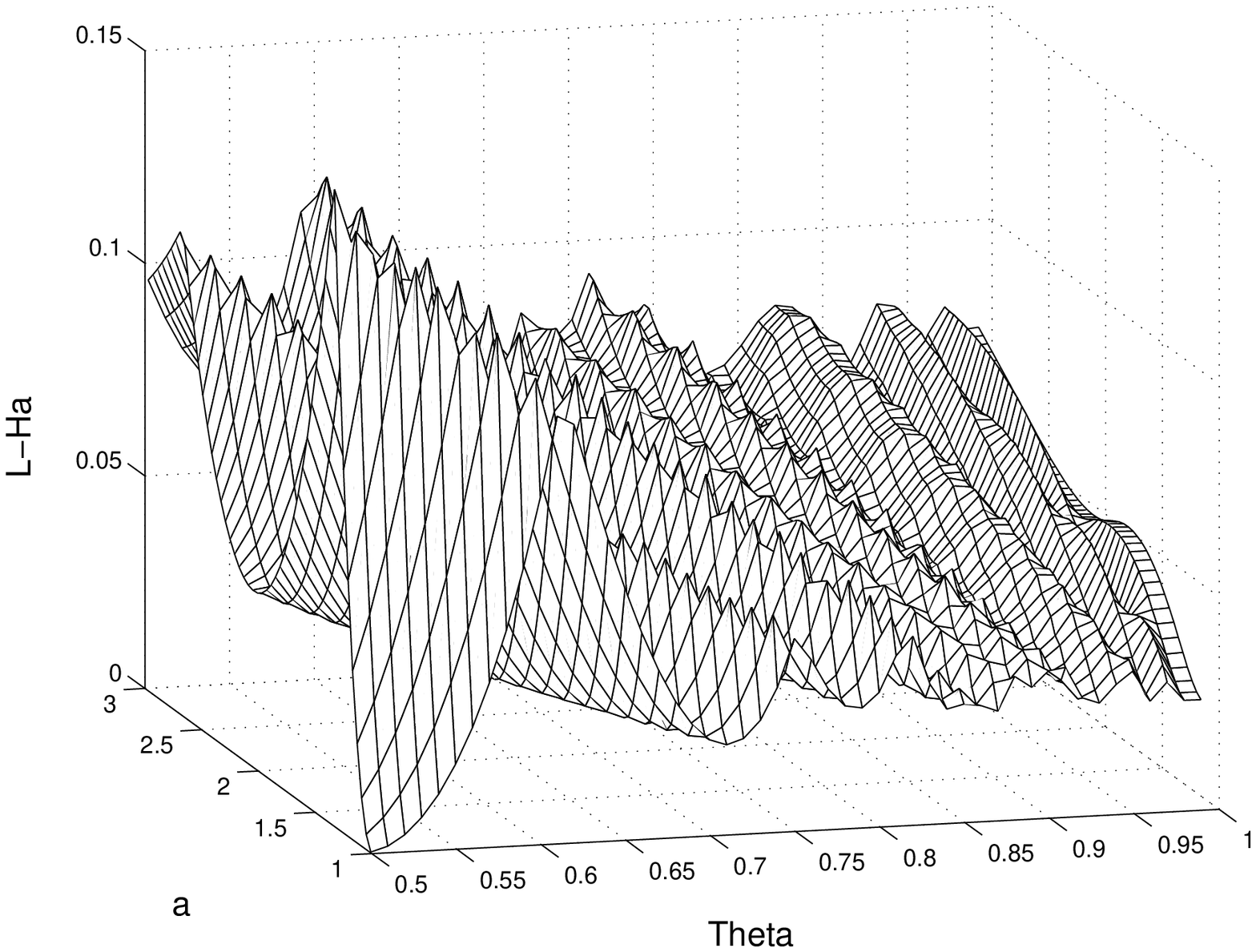} }
     \subfigure[$a<1$]
	       { \label{aneg} \includegraphics[width=.45\textwidth]{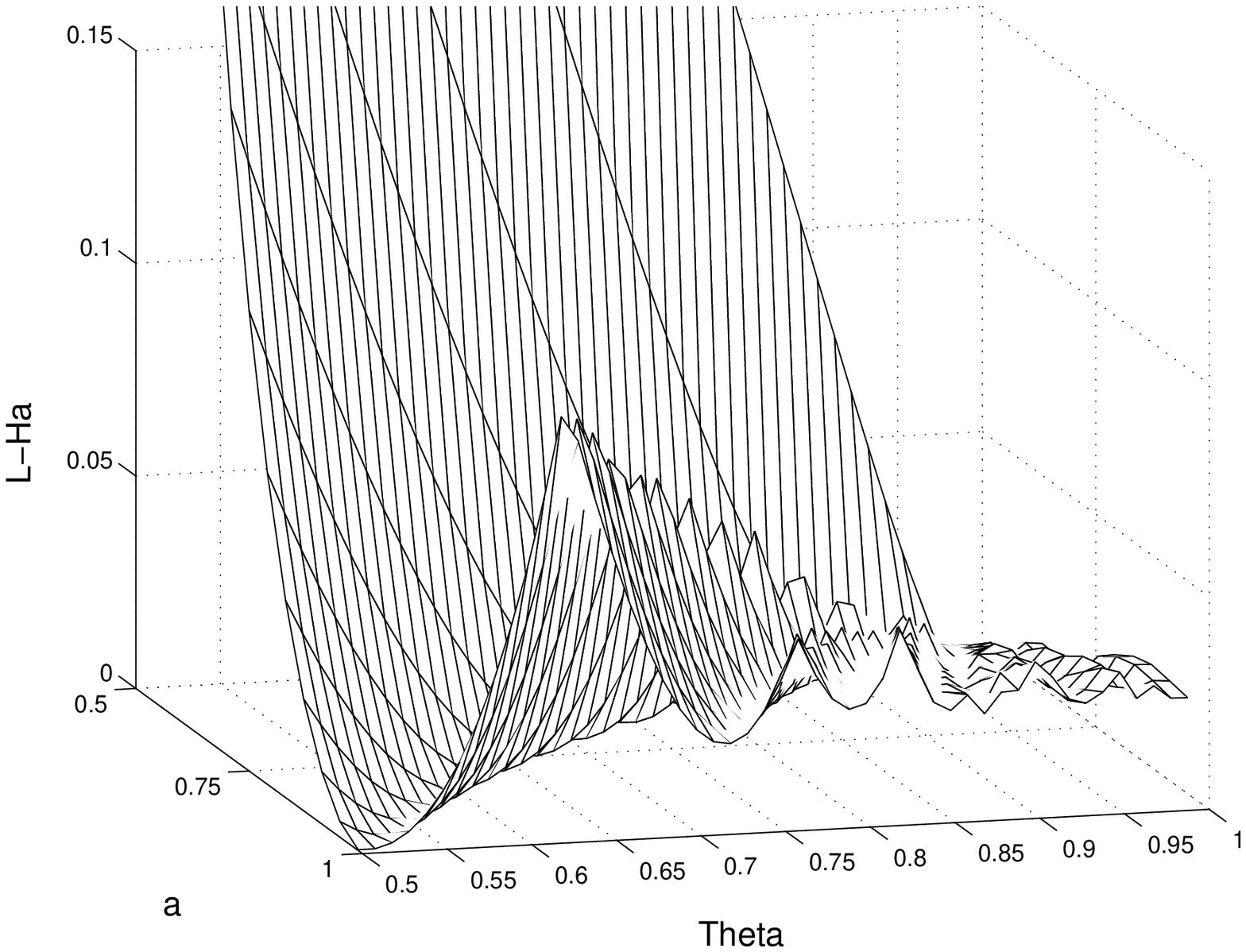} }
     \caption{Redundancy of the optimal code for the geometric
     distribution with the exponential penalty (parameter $a$).
     $\bar{R}_a(\biglen_{\theta,a}^*,\bigp_\theta) =
     \CampCost_a(\bigp_\theta,\biglen_{\theta,a}^*) - H_{\alpha(a)}(\bigp_\theta)$,
     where $\alpha(a) = (1+\lg a)^{-1}$, $\bigp_\theta$ is the
     geometric probability sequence implied by $\theta$, and
     $\biglen_{\theta,a}^*$ is the optimal length sequence for
     distribution $\bigp_\theta$ and parameter $a$.}
     \label{aall}
\end{figure*}

It is equivalent to follow the complete portion of the code with the
unary portion --- as in the exponential Huffman tree in
Fig.~\ref{buildgolo} --- or to reorder the bits and follow the unary
portion with the complete portion --- as in the Golomb
code\cite{Golo}.  The latter is more often used in practice and has
the advantage of being alphabetic, that is, $i>j$ if and only if
$c(i)$ is lexicographically after $c(j)$.

The truncated Golomb tree for any $m \geq \kval-1$ represents a code
that has the same penalty for the $m$-reduced distribution as does the
Golomb code with the corresponding geometric distribution.  We now
show that this is the minimum penalty for any code with this geometric
distribution.

Let $\biglen_{\theta,a}^*$ (or $\biglen^*$ if there is no ambiguity)
be codeword lengths that minimize the penalty for the geometric
distribution (which, as we noted, exist as shown in Theorem~2
of~\cite{Baer06}).  Let $\biglen_m$ be codeword lengths for the
$m$-reduced distribution found earlier; that is, $\len_m(i)$ is the
Golomb length for $i \leq m$ and $\len_m(i) = \len_m(i-\kval)$ for the
remaining values.  Finally, let $\biglen_{\infty}$ be the lengths of
the code implied by $m \rightarrow \infty$, that is, the lengths of
the Golomb code G$\kval$.  Then
\begin{equation}
\begin{array}{rcl}
\displaystyle
\log_a \sum_{i=0}^\infty \boldp(i) a^{\len^*(i)} &\leq&
\displaystyle
\log_a \sum_{i=0}^\infty \boldp(i) a^{\len_{\infty}(i)} \\
&=&
\displaystyle
\log_a \sum_{i=0}^{m+\kval} \boldw_m(i) a^{\len_m(i)} \\
&\leq&
\displaystyle
\log_a \sum_{i=0}^{m+\kval} \boldw_m(i) a^{\len^*(i)} 
\end{array}
\label{fininf}
\end{equation}
where the inequalities are due to the optimality of the respective
codes and the facts that $\boldw_m(i)=\boldp(i)$ for $i \leq m$ and
$$\boldw_m(i)=\sum_{j=0}^\infty (1-\theta)\theta^{i+j\kval}a^{j+1} =
\sum_{j=0}^\infty a^{j+1} \boldp(i+j\kval)$$
for $i \in (m,m+\kval]$.  The difference between the exponent of the
first and the last of the expressions in (\ref{fininf}) is
$$
\begin{array}{l}
\displaystyle
\sum_{i=0}^\infty \boldp(i) a^{\len^*(i)} - \sum_{i=0}^{m+\kval}
\boldw_m(i) a^{\len^*(i)} \\
\displaystyle
\qquad ~ = 
\sum_{i=m+1}^\infty \boldp(i) a^{\len^*(i)}
- \sum_{i=m+1}^{m+\kval} \boldw_m(i) a^{\len^*(i)} .
\end{array}
$$ As $m \rightarrow \infty$ for $m \geq \kval-1$, the sums on the
right-hand side approach~$0$; the first is the difference between a
limit (an infinite sum) and its approaching sequence of finite sums,
all upper bounded in~(\ref{fininf}), and each of the terms in the
second summation is upper-bounded by a multiplicative constant of the
corresponding term in the first.  (In the latter finite summation,
terms are $0$ for $i>m+\kval$.)  Their difference therefore also
approaches zero, so the summations on the left-hand side approach
equality, as do those in (\ref{fininf}), and the Golomb code must be
optimal.
\end{proof}

It is equivalent for the bits of the unary portion to be complemented,
that is, to use $\{0^{\lfloor j/\kval \rfloor} 1 b(j \bmod
\kval,\kval) : j \geq 0\}$ (as in \cite{GaVV}) instead of
$\{1^{\lfloor j/\kval \rfloor} 0 b(j \bmod \kval,\kval) : j \geq 0\}$
(as in \cite{Golo}).  It is also worth noting that Golomb originally
proposed his code in the context of a spy reporting run lengths; this
is similar to R\'{e}nyi's context for communications, related in
Section~\ref{intro} as a motivation for the nonlinear penalty with
$a<1$.

A little algebra reveals that, for a distribution $\bigp_\theta$ and a Golomb
code with parameter $\kval$ (lengths $\biglen_\kval$), 
\begin{equation}
\begin{array}{rcl}
\CampCost_a(\bigp_\theta,\biglen_\kval) &=& \displaystyle
\log_a \sum_{i=0}^\infty
(1-\theta)\theta^i a^{(\left\lceil\frac{i+1-z}{\kval} \right\rceil + g)} \\ 
\displaystyle
&=& g + {\log}_a
\left(1+\frac{(a-1)\theta^z}{1-a\theta^\kval}\right) 
\end{array}
\label{geosum}
\end{equation}
where
$g=\lfloor \log_2 \kval \rfloor + 1$ and $z = 2^g - \kval$.  
Therefore, Theorem~\ref{optgeo} provides the $\kval$ that minimizes
(\ref{geosum}).  If $a>0.5$, the corresponding R\'{e}nyi entropy is
\begin{equation}
H_{\alpha(a)}(\bigp_\theta) = \log_a
\frac{1-\theta}{(1-\theta^{\alpha(a)})^{1/\alpha(a)}}
\label{geoent}
\end{equation}
where we recall that $\alpha(a) = (1 +
\lg a)^{-1}$.  (Again, $a \leq 0.5$ is degenerate, an
optimal code being unary with no corresponding R\'{e}nyi entropy.)

\begin{figure}[t]
\psfrag{L-H}{\mbox{\huge $\bar{R}_1(\biglen_{\theta,1}^*,\bigp_\theta)$}}
\psfrag{THETA}{\mbox{\huge $\theta$}}
     \centering
     \resizebox{8cm}{!}{\includegraphics{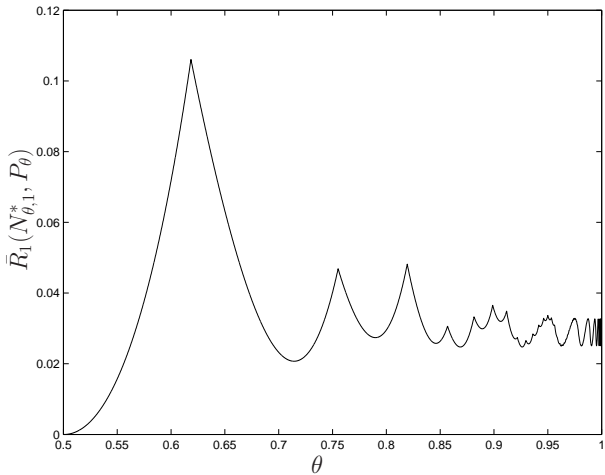}}
     \caption{Redundancy of the optimal code for the geometric
     distribution with the traditional linear penalty.}
     \label{shannon}
\end{figure}

In evaluating the effectiveness of the optimal code, one might use the
following definition of {\defn average pointwise redundancy} (or just
{\defn redundancy}): $$\bar{R}_a(\biglen, \bigp) \definedas
\CampCost_a (\bigp,\biglen) - H_{\alpha(a)}(\bigp) .$$
For nondegenerate values, we can plot the $\bar{R}_a(\biglen_{\theta,a}^*,
\bigp_\theta)$ obtained from the minimization.  This is done for $a>1$
and $a<1$ in Fig.~\ref{aall}.  Note that as $a \rightarrow 1$, the
plot approaches the redundancy plot for the linear case, e.g.,
\cite{GaVV}, reproduced as Fig.~\ref{shannon}.

In many potential applications of nonlinear penalties --- such as the
aforementioned for $a>1$\cite{Jeli,Humb2,BlMc} and $a<1$
(Section~\ref{intro}) --- $a$ is very close to~$1$.  Since the preceding
analysis shows that the Golomb code that is optimal for given $a$
and $\theta$ is optimal not only for these particular values, but for
a range of $a$ (fixing $\theta$) and a range of $\theta$ (fixing $a$),
the Golomb code for the traditional linear penalty is, in some sense,
much more robust and general than previously appreciated.

\section{Other Infinite Sources}
\label{other}

Abrahams noted that, in the linear case, slight deviation from the
geometric distribution in some cases does not change the optimal
code\cite[Proposition~(2)]{Abr1}.  Other extensions to and deviations
of the geometric distribution have also been
considered\cite{MSW,GoMa,BCSV}, including optimal codes for nonbinary
alphabets\cite{Abr1,GoMa}.  Many of these approaches can be adapted to
the nonlinear penalties considered here.  However, in this section we
instead consider another type of probability distribution for binary
coding, the type with a light tail.

Humblet's approach\cite{Humb1}, later extended in \cite{KHN}, uses the
fact that there is an optimal code tree with a unary subtree for any
probability distribution with a relatively light tail, one for which
there is an $r$ such that, for all $j>r$ and $i<j$, $\boldp(i) \geq
\boldp(j)$ and $\boldp(i) \geq \sum_{k=j+1}^\infty \boldp(k)$.  Due to
the additive nature of Huffman coding, items beyond $r$ form the unary
subtree, while the remaining tree can be coded via the Huffman
algorithm.  Once again, this has to be modified for exponential
penalties.

\begin{figure*}
\psfrag{  0x}{\mbox{\tiny $p(0)$}}
\psfrag{  1x}{\mbox{\tiny $p(1)$}}
\psfrag{  2x}{\mbox{\tiny $p(2)$}}
\psfrag{  3x}{\mbox{\tiny $p(3)$}}
\psfrag{  4x}{\mbox{\tiny $p(4)$}}
\psfrag{  5x}{\mbox{\tiny $p(5)$}}
\psfrag{  6x}{\mbox{\tiny $p(6)$}}
\psfrag{  7x}{\mbox{\tiny $p(7)$}}
\psfrag{  8x}{\mbox{\tiny $p(8)$}}
\psfrag{  9x}{\mbox{\tiny $p(9)$}}
\psfrag{  10x}{\mbox{\tiny $p(10)$}}
\psfrag{  11x}{\mbox{\tiny $p(11)$}}
\psfrag{  12x}{\mbox{\tiny $p(12)$}}
\psfrag{  13r}{\mbox{\tiny $w(13)$}}
\begin{center}
\resizebox{14cm}{!}{\includegraphics{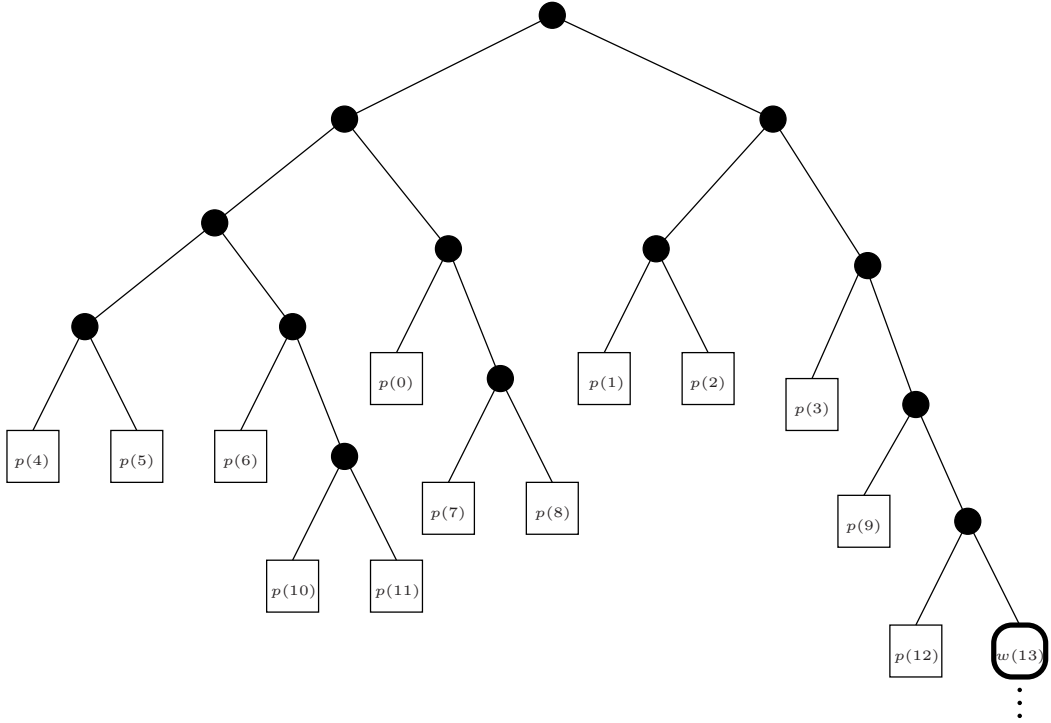}}
\caption{Formation of a unary-ended infinite code using a Huffman-like
code.  (Smaller weights are pictorially lower.)  Weights are merged
bottom-up, in a manner consistent with the exponential Huffman
algorithm, first in the (truncated) unary subtree, then as in the
exponential Huffman algorithm.}
\label{buildhumb}
\end{center}
\end{figure*}

We wish to show that the optimal code can be obtained when there is a
nonnegative integer $r$ such that, for all $j>r$ and $i<j$, $$\boldp(i)
\geq \max\left(\boldp(j), \sum_{k=j+1}^\infty \boldp(k) a^{k-j}\right).$$
The optimal code is obtained by considering the reduced alphabet
consisting of symbols $0,1,\ldots,r+1$ with weights
\begin{equation}
\boldw(i) = \left\{
\begin{array}{ll}
\boldp(i),& i \leq r \\
\sum_{k=r+1}^\infty \boldp(k) a^{k-r},& i = r+1 . \\
\end{array}
\right.
\label{weights}
\end{equation}
Apply exponential Huffman coding to this reduced set of weights.  For
items $0$ through $r$, the Huffman codewords for the reduced and the
infinite alphabets are identical.  Each other item $i>r$ has a
codeword consisting of the reduced codeword for item $r+1$ (which,
without loss of generality, consists of all $1\s$) followed by the
unary code for $i-r-1$, that is, $i-r-1$ ones followed by a zero.  We
call such codes {\defn unary-ended}.  A pictorial example is shown in
Fig.~\ref{buildhumb} for a problem instance for which $r=12$.

\begin{theorem}
Let $\boldp(\cdot)$ be a probability measure on the set of nonnegative
integers, and let $a$ be the parameter of the penalty to be optimized.
If there is a nonnegative integer $r$
such that for all $j>r$ and $i<j$,
\begin{equation}
\boldp(i) \geq \boldp(j)
\label{cond1}
\end{equation}
and
\begin{equation}
\boldp(i) \geq \sum_{k=j+1}^\infty \boldp(k) a^{k-j}
\label{cond2}
\end{equation}
then there exists a minimum-penalty binary prefix code with every
codeword~$j>r$ consisting of $j-x$ $1\s$ followed by one $0$ for some
fixed nonnegative integer~$x$.
\label{tailthm}
\end{theorem}

\begin{proof}
The idea here is similar to that for geometric distributions, to show
a sequence of finite codes which in some sense converges to the
optimal code for the infinite alphabet.  In this case we consider the
infinite sequence of codes implicit in the above; for a given $m \geq -1$, the
corresponding codeword weights are
$$
\boldw_m(i) = \left\{
\begin{array}{ll}
\boldp(i),& i < r+m+2 \\
\sum_{k=r+m+2}^\infty \boldp(k) a^{k-r-m-1},& i = r+m+2. \\
\end{array}
\right.
$$  It is obvious that an optimal code for
each $m$-reduced distribution is identical to the proposed code for
the infinite alphabet, except for the item $r+m+2$, which is the
code tree sibling of item $r+m+1$.  

For $a<1$, we show, as in the geometric case, that the difference
between the penalties for the optimal and the proposed codes
approaches~$0$.  In this case, the equivalent of
inequality~(\ref{fininf}) is
\begin{equation}
\begin{array}{rcl}
\displaystyle
\log_a \sum_{i=0}^\infty \boldp(i) a^{\len^*(i)} &\leq&
\displaystyle
\log_a \sum_{i=0}^\infty \boldp(i) a^{\len_{\infty}(i)} \\
&=&
\displaystyle
\log_a \sum_{i=0}^{r+m+2} \boldw_m(i) a^{\len_m(i)} \\
&\leq&
\displaystyle
\log_a \sum_{i=0}^{r+m+2} \boldw_m(i) a^{\len^*(i)} 
\end{array}
\label{fininf2}
\end{equation}
where in this case $n_\infty(i)$ denotes a codeword of the proposed
code, $n_m(i) = n_\infty(i)$ for $i<r+m+2$ and $n_m(i) =
n_\infty(i-1)$ for $i=r+m+2$, and, again, $\len^*(\cdot)$ denotes the
lengths of codewords in an optimal code.  The corresponding difference
between the exponent of the first and the last expressions of
(\ref{fininf2}) is
\begin{equation}
\begin{array}{l}
\displaystyle
\sum_{i=0}^\infty \boldp(i) a^{\len^*(i)} - 
\sum_{i=0}^{r+m+2} \boldw_m(i) a^{\len^*(i)} \\
\displaystyle
\qquad = 
\sum_{i=r+m+2}^\infty 
\boldp(i) a^{\len^*(i)} - \boldw_m(r+m+2) 
a^{\len^*(r+m+2)}. 
\end{array}
\label{fininf3}
\end{equation}
As $m \rightarrow \infty$, both terms in the difference
on the second line of (\ref{fininf3}) clearly approach $0$, so the
terms in~(\ref{fininf2}) approach equality, showing the proposed code to
be optimal.

For $a>1$, the same method will work, but it is not so obvious that
the terms in the difference on the second line of (\ref{fininf3})
approach~$0$.  Let us first find an upper bound for $\boldw_m(r+m+2)$
in terms of $\boldp(r+m+2)$:
\begin{eqnarray*}
\boldw_m(r+m+2) 
&=& a\boldp(r+m+2)+a^2\boldp(r+m+3)+\\
&& \displaystyle\qquad \sum_{i=r+m+4}^\infty \boldp(i) a^{i-r-m-1} \\
&\leq& (a^2+a)\boldp(r+m+2)+a^2\boldp(r+m+3) \\
&\leq& (2a^2+a)\boldp(r+m+2)
\end{eqnarray*}
where the first equality is due to the definition of
$\boldw_m(\cdot)$, the first inequality due to (\ref{cond2}), and the
second inequality due to (\ref{cond1}).  Thus $\boldw_m(r+m+2)$ has an
upper bound of $(2a^2+a)\boldp(r+m+2)$ for all $m \geq -1$.  In
addition, since the proposed code has a finite penalty --- identical
to that of any reduced code --- the optimal code has a finite penalty,
and the sequence of its terms --- each one of which has the form
$\boldp(r+m+2) a^{\len^*(r+m+2)}$ --- approaches $0$ as $m$ increases.
Thus $\boldw_m(r+m+2) a^{\len^*(r+m+2)}$ approaches $0$ as well.  Due
to the optimality of $\len^*(\cdot)$, $\boldw_m(r+m+2)
a^{\len^*(r+m+2)}$ serves as an upper bound for $\sum_{i=r+m+2}^\infty
\boldp(i) a^{\len^*(i)}$, and thus both terms approach~$0$.  As with
$a<1$, then, the terms in~(\ref{fininf2}) approach equality for $m
\rightarrow \infty$, showing the proposed code to be optimal.
\end{proof}

\begin{figure}[t]
\psfrag{a}{$a$}
\psfrag{ag}{\mbox {\huge $a$}}
\psfrag{g}{\mbox {\huge $g$}}
     \centering
     \resizebox{8cm}{!}{\includegraphics{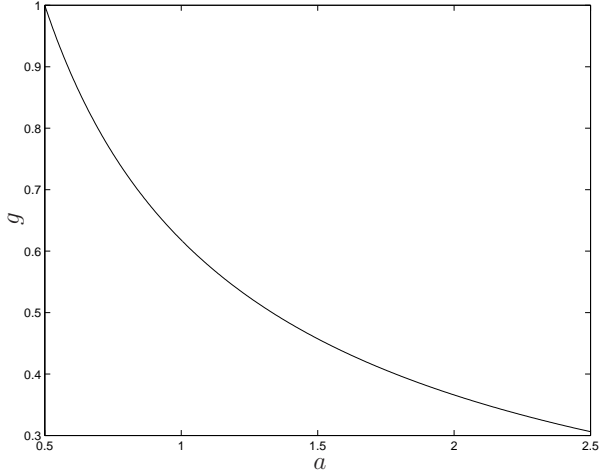}}
     \caption{Ratio $g$, probability distribution fall-off sufficient
     for the optimality of a unary-ended code.  Note that 
     $1/g = \Phi \definedas \frac{1}{2}(1+\sqrt{5})$,
     the golden ratio, at $a=1$.}
     \label{ga}
\end{figure}

The rate at which $\boldp(\cdot)$ must decrease in order to satisfy
condition~(\ref{cond2}) clearly depends on $a$.  One simple sufficient
condition --- provable via induction --- is that it satisfy $\boldp(i)
\geq a \boldp(i+1) + a \boldp(i+2)$ for large $i$.  A less general
condition is that $\boldp(i)$ eventually decrease at least as quickly
as $g^i$ where $g = (\sqrt{1+4/a}-1)/2$, the same ratio needed for a
unary geometric code for $\theta=g$, as in~(\ref{ineq}).  The ratio
$g$ is plotted in Fig.~\ref{ga}.

For $a \rightarrow 1$, these conditions approach those derived in
\cite{Humb1}.  The stronger results of \cite{KHN} do not easily extend
here due to the nonadditivity of the exponential penalty.  An attempt
at such an extension in \cite[pp.~103--105]{Baer} gives no criteria
for success, so that, while one could produce certain codewords for
certain codes, one might fail in producing other codewords for the
same codes or for other codes.  Thus this extension is not truly a
workable algorithm.

Consider the example of optimal codes for the Poisson distribution,
$$\boldp_\lambda(i)=\frac{\lambda^i e^{-\lambda}}{i!} . $$ How does
one find a suitable value for $r$ (as in Section~\ref{other}) in such
a case?  It has been shown that $r \geq \lceil e \lambda \rceil - 1$
yields $\boldp(i) \geq \boldp(j)$ for all $j>r$ and $i<j$, satisfying
the first condition of Theorem~\ref{tailthm} \cite{Humb1}.  Moreover,
if, in addition, $j \geq \lceil 2 a \lambda \rceil - 1$ (and thus $j >
a \lambda - 1$), then
\begin{eqnarray*}
\sum_{k=1}^\infty \boldp(j+k)a^k 
&=& \frac{e^{-\lambda}\lambda^j}{j!}\left[
\frac{a \lambda}{j+1} + \frac{a^2 \lambda^2}{(j+1)(j+2)} + \cdots \right] \\
&<& \boldp(j) \left[\frac{a \lambda}{j+1} + \frac{a^2 \lambda^2}{(j+1)^2} + \cdots \right] \\
&=& \boldp(j) \frac{\frac{a \lambda}{j+1}}{1-\frac{a \lambda}{j+1}} \\
&\leq& \boldp(j) \\
&\leq& \boldp(i) .
\end{eqnarray*}
Thus, since we consider $j > r$, $r = \max(\lceil 2 a \lambda \rceil -
2, \lceil e \lambda \rceil - 1)$ is sufficient to establish an $r$
such that the above method yields the optimal infinite-alphabet code.

In order to find the optimal reduced code, use
$$\boldw_{-1}(r+1)=\sum_{k=r+1}^\infty \boldp(k) a^{k-r} = a^{-r}e^{\lambda(a-1)} - \sum_{k=0}^r \boldp(k) a^{k-r} .
$$  For example, consider the Poisson distribution with $\lambda = 1$.  We
code this for both $a=1$ and $a=2$.  For both values, $r = 2$, so both
are easy to code.  For $a=1$, $\boldw_{-1}(3) = 1 - 2.5 e^{-1} \approx
0.0803 \ldots$, while, for $a=2$, $\boldw_{-1}(3) = 0.25 e - 1.25
e^{-1} \approx 0.2197 \ldots$.  After using the appropriate Huffman
procedure on each reduced source of $4$ weights, we find that the
optimal code for $a=1$ has lengths $\biglen = \{1, 2, 3, 4, 5, 6, \ldots\}$
--- those of the unary code --- while the optimal code for $a=2$ has
lengths $\biglen = \{2, 2, 2, 3, 4, 5, \ldots\}$.

It is worthwhile to note that these techniques are easily extensible
to finding an optimal alphabetic code --- that is, one with $c(i)\s$
arranged in lexicographical order --- for $a>1$.  One needs only to
find the optimal alphabetic code for the reduced code with weights
given in equation~(\ref{weights}), as in \cite{HKT}, with codewords
for $i>r$ consisting of the reduced code's codeword for $r+1$ followed
by $i-r-1$ ones and one zero.  As previously mentioned, Golomb codes
are also alphabetic and thus are optimal alphabetic codes for the
geometric distribution.

\section{Application: Buffer Overflow}
\label{application}

The application of the exponential penalty in \cite{Humb2} concerns
minimizing the probability of a buffer overflowing.  It requires that
each candidate code for overall optimality be an optimal code
for one of a series of exponential parameters ($a\s$ where $a>1$).  An
iterative approach yields a final output code by noting that, for the
overall utility function, each candidate code is no worse than 
its predecessor, and there are a finite number of possible candidate
codes.  Therefore, eventually a candidate code yields the same value
as the prior candidate code, and this can be shown to be the optimal
code.  This application of exponential Huffman coding can, using the
above techniques, be extended to infinite alphabets.

In the application, integers with a known distribution $\bigp$ arrive
with independent intermission times having a known probability density
function.  Encoded bits are sent at a given rate, with bits to be sent
waiting in a buffer of fixed size.  Constant $b$ represents the buffer
size in bits, random variable $T$ represents the probability
distribution of source integer intermission times measured in units of
encoded bit transmission time, and function $A(s)$ is the
Laplace-Stieltjes transform of $T$, $\E[e^{-sT}]$.  

When the integers are coded using $\biglen = \{\len(i)\}$, the
probability per input integer of buffer overflow is of the order of
$e^{-s^*b}$, where $s^*$ is the largest $s$ such that
$$
f(\biglen,s) \leq 1
$$
where
\begin{equation}
f(\biglen,s) \definedas A(s) \sum_{i=0}^\infty
\boldp(i)e^{s\len(i)} .
\label{buff}
\end{equation}

The previously known algorithm to maximize $s^*$ is as follows:

{\bf Procedure for Finding Code with Largest $s^*$} \cite{Humb2}

\begin{enumerate}
\item Choose any $s_0 \in \Rp$.
\item $j \leftarrow 0$.
\item $j \leftarrow j+1$.
\item Find codeword lengths $\biglen_j$ minimizing $\sum_i \boldp(i) e^{s_{j-1}
\len(i)}$.
\item Compute $s_j \definedas \max\{s \in \R : f(\biglen_j,s) \leq 1\}$.
\item If $s_j \neq s_{j-1}$ then go to step 3; otherwise stop.
\end{enumerate}

We can use the above methods in order to accomplish step 4, but we
still need to examine how to modify steps 1 and 5 for an infinite
input alphabet.

First note that, unlike in the finite case, $s^*<\infty$, that is, there
always exists an $s^* \in \Rp$ such that, for all $s>s^*$,
$f(\biglen,s) > 1$.  For any stable system, the buffer cannot receive
integers more quickly than it can transit bits, so there is a positive
probability that $\P[T \geq 1]$.  Thus the Laplace-Stieltjes transform
$A(s)$ exceeds $c_1 e^{-s}$ for some constant $c_1>0$.  Also, without
loss of generality, we can assume that $\boldp(i)$ is monotonic
nonincreasing and an optimal $\len(i)$ is monotonic nondecreasing.
This monotonicity means that $\len(i) \geq \lg i$, and there is no
exponential base $a_0$ and offset constant $c_2$ for which
$\sum_{i=0}^\infty \boldp(i) e^{s\len(i)} \leq a_0^{s+c_2}$ for all~$s
\in \Rp$.  Thus the summation in~(\ref{buff}) must increase
superexponentially, and, multiplying the $A(s)$ and summation terms,
there is an $s$ such that $f(\biglen,s)>1$ for $s>s^*$.

For step 1, the initial guess proposed in \cite{Humb2} is an upper
bound for all possible values of $s^*$.  The R\'{e}nyi entropy of
$\bigp$ is used to find an initial guess using
\begin{equation}
A(s) \left(\sum_{i=0}^\infty \boldp(i)^{\frac{1}{1+\lg e^s}}\right)^{1+\lg e^s}
\leq A(s) \sum_{i=0}^\infty \boldp(i)e^{s\len(i)},
\label{humbbound}
\end{equation}
and choosing $s_0$ as the largest $s$ such that the left term of
(\ref{humbbound}) is no greater than one.  Thus, $s_0 \geq s^*$ for any
value of $s^*$ corresponding to step 5.

This technique is well-suited to a geometric distribution, for which
entropy has the closed form shown in equation (\ref{geoent}), so
$$A(s) \cdot \frac{1-\theta}{\left(1-\theta^{(1+\lg
e^s)^{-1}}\right)^{1+\lg e^s}} \leq f(\biglen,s).$$ However, a general
distribution with a light tail, such as the Poisson distribution,
might have no closed form for this bound.  One solution to this is to
use more relaxed lower bounds on the sum --- such as using a partial
sum with a fixed number of terms --- yielding looser upper bounds
for~$s^*$.  Another approach would be to note that, because of the
light tail, the infinite sum can usually be quickly calculated to the
precision of the architecture used.  Note, however, that no matter
what the technique, the bound must be chosen so that $s_0$ is an
real number and not infinity.  Partial sums may be refined to accomplish
this.

In calculating $f(\biglen,s)$ for use in step 5, the geometric
distribution has the closed-form value for $f$ obtainable from
equation (\ref{geosum}), while the other distributions must instead
rely on approximations of~$f$.  As before, this is easily done due to
the light tail of the distribution.  Alternatively, a partial sum and
a geometric approximation can be used to bound $f(\biglen,s)$ and thus
$s^*$, and these two bounds used to find two codes.  If the two codes
are identical, the algorithm may proceed; otherwise, we must roll back
to the summation and improve the bounds until the codes are identical.

These variations make the steps of the algorithm possible, but the
algorithm itself must also be proven correct with the variations.

\begin{theorem}
Given a geometric distribution or an input distribution satisfying the
conditions of Theorem~\ref{tailthm} for $a=e^{s_0}$, where $s_0$ is an
upper-bound on $s^*$, the above Procedure for Finding Code
with Largest $s^*$ terminates with an optimal code.
\label{qthm}
\end{theorem}

\begin{proof}
The number of codes that can be generated in the course of running the
algorithm should be bounded so that the algorithm is guaranteed to
terminate.  Optimality for the algorithm then follows as for the
finite case~\cite{Humb2}.  As in the finite case, $s_{j+1} \geq s_j$
for $j \geq 1$ (but not $j=0$) due to step 5 [$f(\biglen_j,s_j) \leq
1$], step 4 [$f(\biglen_{j+1},s_j) \leq f(\biglen_j,s_j)$], and the
definition of $s_{j+1}$.  

In the case of a geometric distribution,
each $\biglen_j$ is a Golomb code G$\kval_j$ for some positive
integer~$\kval_j$.  Clearly, if we choose $s_0$ as detailed above, it
is the greatest value of $s_j$, being either optimal or unachievable
due to its derivation as a bound of the problem.  Since
$\mbox{G}\kval_i$ (with lengths $\biglen_i$) is the optimal code for
the code with exponential base $a=e^{s_{i-1}}$, (\ref{ineq}) means
that $\theta^{\kval_i} + \theta^{\kval_i+1} \leq e^{-s_{i-1}} <
\theta^{\kval_i-1} + \theta^{\kval_i}$, and thus
$$(1+\theta)\theta^{\kval_1} \leq e^{-s_0} \leq e^{-s_{j-1}} <
(1+\theta)\theta^{\kval_j-1}$$ and, since $\theta < 1$, we have
$\kval_j-1 < \kval_1$ (or, equivalently, $\kval_j \leq \kval_1$) for all
$j \geq 1$.  Therefore, there are only $\kval_1$ possible codes the
algorithm can generate.  

In the case of a distribution with a lighter tail, the minimum $r$ of
Theorem~\ref{tailthm} increases with each iteration after the first,
and the first $r_1$ (corresponding to $s_0$) upper bounds the
remaining $r_i$.  Thus all candidate codes can be specified by their
first $r_1$ codeword lengths, none of which is greater than $r_1$.
The number of codes is then bounded for both cases, and the algorithm
terminates with the optimal code.
\end{proof}

\section{Redundancy penalties}
\label{nonexp}

It is natural to ask whether the above results can be extended to
other penalties.  One penalty discussed in the literature is that of
maximal pointwise redundancy\cite{DrSz}, which is
$$R^*(\biglen,\bigp) \definedas \sup_{i \in \X} [\len(i)+\lg
\boldp(i)]$$ where we use $\sup$ when we are not assured the existence
of a maximum.  This can be shown to be a limit of the exponential
case, as in \cite{Baer05}, allowing us to analyze its minimization
using the same techniques as exponential Huffman coding.  This limit
can be shown by defining {\defn $d$th exponential redundancy} as
follows:
\begin{eqnarray*}
R_d(\biglen,\bigp) &\definedas&
\frac{1}{d} \lg \sum_{i \in \X} \boldp(i) 2^{d\left(\len(i)+\lg \boldp(i)\right)} \\
 &=& \frac{1}{d} \lg
\sum_{i \in \X} \boldp(i)^{1+d} 2^{d\len(i)}.
\end{eqnarray*}
Thus $R^*(\biglen,\bigp) = \lim_{d \rightarrow \infty}
R_d(\biglen,\bigp)$, and the above methods should apply in the limit.
In particular:

\begin{theorem}
The Golomb code G$\kval$ for $\kval = \lceil -1/\lg \theta \rceil$ 
is optimal for minimizing maximal pointwise redundancy for $\bigp_\theta$. 
\label{optmmr}
\end{theorem}

\begin{figure*}
\psfrag{DABRRR}{$R_d(\biglen_{\theta,a,d}^*,\bigp_\theta)$}
\psfrag{Theta}{$\theta$}
\psfrag{theta}{$\theta$}
\psfrag{THETA}{\mbox{\huge $\theta$}}
     \centering
\begin{picture}(0,0)%
\includegraphics{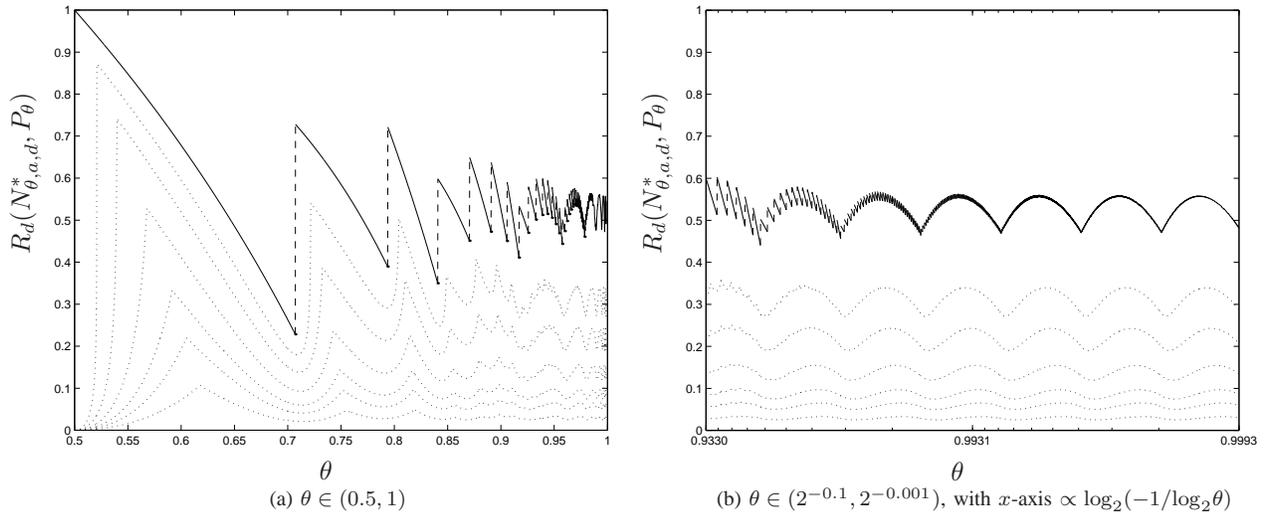}%
\end{picture}%
\setlength{\unitlength}{1865sp}%
\begingroup\makeatletter\ifx\SetFigFont\undefined%
\gdef\SetFigFont#1#2#3#4#5{%
  \reset@font\fontsize{#1}{#2pt}%
  \fontfamily{#3}\fontseries{#4}\fontshape{#5}%
  \selectfont}%
\fi\endgroup%
\begin{picture}(16332,6837)(-8,-6007)
\put(4051,-5911){\makebox(0,0)[b]{\smash{{\SetFigFont{8}{9.6}{\familydefault}{\mddefault}{\updefault}{(a) $\theta \in (0.5,1)$}%
}}}}
\put(12511,-5911){\makebox(0,0)[b]{\smash{{\SetFigFont{8}{9.6}{\familydefault}{\mddefault}{\updefault}{(b) $\theta \in (2^{-0.1},2^{-0.001})$, with $x$-axis $\propto \lg (- 1/{\lg \theta})$}%
}}}}
\end{picture}%
     \caption{Maximal pointwise redundancy of the optimal maximal
     redundancy code for the geometric distribution, solid
     (with discontinuities represented by dashed); optimal $d$th exponential
     redundancy for the geometric distribution, dotted for
     $d=\{1,2,4,16,256,65536\}$, from lowest to highest.}
     \label{mmr}
\end{figure*}

\begin{proof}

{\it Case 1:} Consider first when $-1/\lg \theta$ is not an integer.
We show that $\kval = \lceil -1/\lg \theta\rceil$ is optimal by
finding a $D$ such that, for all $d > D$, the optimal code for the
$d$th exponential redundancy penalty is G$\kval$.  For
a fixed $d$, (\ref{ineq}) implies that such a code should satisfy
\begin{equation}
(\theta^{1+d})^\kval + (\theta^{1+d})^{\kval+1} \leq \frac{1}{2^d} <
(\theta^{1+d})^{\kval-1} + (\theta^{1+d})^\kval,
\label{dineq}
\end{equation}
and thus we wish to show that this holds for all $d > D$.  Consider
$\kvals = \lceil -1/\lg \theta\rceil$.  Clearly, $\kvals >
-1/\lg \theta$, or, equivalently, 
\begin{equation}
\theta^\kvals < \frac{1}{2}.
\label{mmr1}
\end{equation}
Now consider $$D=-1+\frac{1}{1+(\kvals-1)\lg \theta}$$ so that
$(\kvals-1)\lg \theta \in (-1,0]$ and therefore $D \geq 0.$ Taken
together with the fact that $\theta \in (0,1)$, (\ref{mmr1}) yields
$\theta^{d\kval} < 2^{-d}$ and $(1+\theta^{1+d})\theta^\kval <
2\theta^k < 1$.  Multiplication yields the left-hand side of
(\ref{dineq}) for any $d > D$.  For any such $d$, algebra easily shows
that we also have the inequality $(2\theta^{\kvals-1})^{1+d} \geq 2$,
yielding
\begin{eqnarray*}
\left[(\theta^{1+d})^{\kvals-1}+(\theta^{1+d})^{\kvals}\right]2^d 
&=& \frac{1}{2}(2\theta^{\kvals-1})^{1+d} + 
\frac{1}{2}(2\theta^{\kvals})^{1+d} \\
&=& \frac{1}{2}(2\theta^{\kvals})^{1+d}
(\theta^{-1-d}+1) \\
&=& \frac{1}{2}(2\theta^{\kvals-1})^{1+d}
(1+\theta^{1+d}) \\
&>& 1 .
\end{eqnarray*}
This is equivalent to the right-hand side of
inequality~(\ref{dineq}) for the values implied by the definition of
$R_d(\biglen,\bigp)$.  Then G$\kvals$ is an optimal code for
$d > D$, and thus for the limit case of maximal pointwise redundancy.

{\it Case 2:} Now consider when $-1/\lg \theta$ is an integer.  It
should be noted that, for the traditional (linear) penalty, these are
precisely the $\kval$ values that Golomb considered in his original
paper\cite{Golo} and that they are local infima for the minimum
maximal pointwise redundancy function in~$\theta$, as in
Fig.~\ref{mmr}.  Here we show they are local minima.

Since $\theta=0.5$ is a dyadic probability distribution and thus
trivial, we can assume that $\theta > 0.5$.  We wish to show that
optimality is preserved in these right limits of Case~1.  Note that,
for each $i$ with fixed $\biglen$,
$$\lim_{\theta' \uparrow \theta} \left[\len(i) + \lg
\boldp_{\theta'}(i) \right] = \len(i) + \lg \boldp_{\theta}(i).$$ This
is of particular interest for the value of $i$ maximizing pointwise
redundancy for G$\kval$ at $\theta'$, where $\theta' \in
(\theta^{1/\lg 2\theta}, \theta)$, allowing us to use the right limit
of $\theta$.  Let $i^{**} \definedas 2^{\lceil \lg \kval \rceil}-\kval$, the
smallest $i$ which has codeword length exceeding the codeword length
for item~$0$.  Clearly the pointwise redundancy for this value is
greater than that for all items with $i<i^{**}$, since they are one
bit shorter but not more than twice as likely.  Similarly, items in
$(i^{**},\kval)$ have identical length but lower probability, and thus
smaller redundancy.  For items with $i \geq \kval$, note that the
redundancy of items in the sequence $\{j, j+\kval, j+2\kval, \ldots\}$
for any $j$ must be nonincreasing because the difference in redundancy
is constant yet redundancy is upper-bounded by the maximum.  Thus
$i^{**}$ maximizes pointwise redundancy for G$\kval$ at $\theta'$.

We know the pointwise redundancy of $i^{**}$ for G$\kval$ 
at $\theta$, although we have yet to show that $i^{**}$ yields the
maximal pointwise redundancy for G$\kval$ at $\theta$ or that G$\kval$ 
minimizes maximal pointwise redundancy.  However, for any code,
including the optimal code, as a result of pointwise continuity,
\begin{eqnarray*}  
\sup_{i \in \X_\infty} [\len(i)+\lg \boldp_\theta(i)] &\geq& \len(i^{**}) + \lg
\boldp_\theta(i^{**}) \\ &=& \lim_{\theta' \uparrow \theta} [\len(i^{**}) +
\lg \boldp_{\theta'}(i^{**})] .
\end{eqnarray*}
From the above discussion, it is clear that the right-hand side is
minimized by the Golomb code with $\kval=-1/\lg \theta$, so, because
the left-hand side achieves same value with this code, the left-hand
side is indeed minimized by G$\kval$.  Thus this code
minimizes maximal pointwise redundancy for~$\theta$.  The
corresponding maximal pointwise redundancy is
$$ 
\begin{array}{l}
\max_i [\len_\theta^{**}(i)+\lg \boldp_\theta(i)] \\
\begin{array}{rcl}
&=& \len_\theta^{**}(2^{\lceil \lg \kval \rceil}-\kval) +\lg
\boldp_\theta(2^{\lceil \lg \kval \rceil}-\kval) \\ 
&=& \lceil
\lg \kval \rceil + 1 + \lg(1-\theta) + (2^{\lceil \lg \kval
\rceil}-\kval) \lg \theta 
\end{array}
\end{array}
$$
where $\biglen_\theta^{**} = \{\len_\theta^{**}(i)\}$ is defined as
the lengths of a code minimizing maximal pointwise redundancy.  Note
that this is the redundancy for all items $i=2^{\lceil \lg \kval
\rceil}+ j \kval$ with integer $j \geq -1$.
\end{proof}

It is worthwhile to observe the behavior of maximal pointwise
redundancy in a fixed (not necessarily optimal) Golomb code with
length distribution $\biglen_\kval$.  The maximal pointwise redundancy
$$R^*(\biglen_\kval,\bigp_\theta) = \sup_{i \in \X_\infty}
[\len_\kval(i)+\lg \boldp_\theta(i)]$$ decreases with increasing
$\theta$ --- and is an optimal code for $\theta \in (2^{-1/(\kval-1)},
2^{-1/\kval}]$ --- until $\theta$ exceeds $2^{-1/\kval}$, after which
there is no maximum, that is, pointwise redundancy is unbounded.  This
explains the discontinuous behavior of minimum maximal redundancy for
an optimal code as a function of $\theta$, illustrated in
Fig.~\ref{mmr}, where each continuous segment corresponds to an
optimal code for $\theta \in (2^{-1/(\kval-1)}, 2^{-1/\kval}]$.

Note also the oscillating behavior as $\theta \uparrow 1$.  We show in
Appendix~\ref{maxred} that $\lim \inf_{\theta \uparrow 1}
R^*(\biglen_\theta^{**},\bigp_\theta) = 1-\lg \lg e$ and $\lim
\sup_{\theta \uparrow 1} R^*(\biglen_\theta^{**},\bigp_\theta) = 2 -
\lg e$, and we characterize this oscillating behavior.  This technique
is extensible to other redundancy scenarios of the kind introduced
in~\cite{Baer05}.

For distributions with light tails, one can use a technique much like
the technique of Theorem~\ref{tailthm} in Section~\ref{other}.  First
note that this requires, as a necessary step, the ability to construct
a minimum maximal pointwise redundancy code for finite alphabets.
This can be done either with the method in \cite{DrSz} or any of those
in \cite{Baer05}, the simplest of which uses a variant of the
tree-height problem\cite{Park}, solved via a different extension of
Huffman coding.  Simply put, the weight combining rule, rather than
$\boldw(j) + \boldw(k)$ or $a \cdot (\boldw(j) + \boldw(k))$, is
\begin{equation}
\tilde{\boldw}(j) = 2\cdot\max(\boldw(j),\boldw(k)).
\label{maxrule}
\end{equation}
This rule is used to create an optimal code with lengths
$\biglen^{(r)}$ for $\bigw^{(r)} \definedas \{\boldp(0), \boldp(1),
\ldots, \boldp(r), 2\boldp(r+1)\}$, assuming a unary subtree for items
with index $i\geq r$ (and no other items) is part of an optimal code
tree.  As in the coding method corresponding to Theorem~\ref{tailthm},
the codewords for items $0$ through $r$ of this reduced code are
identical to those of the infinite alphabet.  Each other item $i>r$
has a codeword consisting of the reduced codeword for $r+1$ followed
by the unary code for $i-r-1$, that is, $i-r-1$ ones followed by a
zero.

A sufficient condition for using this method is finding an $r$ such that
$$\mbox{for all } i<r,~\boldp(i) \geq \boldp(r)$$
and 
$$\mbox{for all } j \geq r,~\boldp(j) \geq 2 \boldp(j+1).$$ For
such~$j$, pointwise redundancy is nonincreasing along a unary subtree,
as 
\begin{eqnarray*}
\len(j) + \lg \boldp(j) &=& \len(j+1) + \lg (\boldp(j)/2) \\
&\geq& \len(j+1) + \lg \boldp(j+1).
\end{eqnarray*}
The aforementioned coding method works because, for each $j$, an
optimal subtree consisting of the items with index $i\geq j$ and
higher has $\len(i) = \len(j) - j + i$; this subtree is optimal because the
weight of the root node of {\it any} subtree cannot be less than
$2\boldp(j)$.  A formal proof, similar to that of
Theorem~\ref{tailthm}, is omitted in the interest of space.

For a Poisson random variable, $r = \lceil e \lambda \rceil - 1$
satisfies this condition, since, for $i < r \leq j$, $\boldp(i) \geq
\boldp(r)$ (as in \cite{Humb1}), and
$$\boldp(j) = \frac{j+1}{\lambda} \boldp(j+1) \geq \frac{r+1}{\lambda}
\boldp(j+1) \geq e\boldp(j+1) > 2\boldp(j+1).$$  Thus such a random
variable can be coded in this manner.

Note that other sufficient conditions can be obtained through
alternative methods.  One simple rule is that any code for which $p(i)
\leq 2^{-i}p(0)$ for all $i > 0$ will necessarily have $\len(0) + \lg
p(0)$ minimized by letting $\len(0)=1$, and this will be the maximum
redundancy if $\len(i)=i-1$ in general.  For example, a unary tree
optimizes $\bigp = \{0.6, 0.15, 0.15, 0.0375, 0.0375, \ldots\}$, since
$\lg 1.2 \approx 0.263$ is a lower bound on maximal pointwise
redundancy for any code given $p(1)=0.6$, and this bound is achieved
for the unary code.  If viewed as a rule for a unary subtree, this is
looser than the above, since, unlike linear and exponential penalties,
not all subtrees of the subtree need be optimal.  Other relaxations
can be obtained, although, as they are usually not needed, we do not
discuss them here.

\section{Conclusion}

The aforementioned methods for coding integers are applicable to
geometric and light-tailed distributions with exponential and related
penalties.  Although they are not direct applications of Huffman
coding, per se, these methods are derived from the properties of
generalizations of the Huffman algorithm.  This allows examination of
subtrees of a proposed optimal code independently of the rest of the
code tree, and thus specification of finite codes which in some sense
converge to the optimal integer code.  Different penalties --- e.g.,
$\varphi(x) = x^2$, implying the minimization of $\sqrt{\sum_i
\boldp(i) \len(i)^2}$ --- do not share this independence property, as
an optimal code tree with optimal subtrees need not exist.  Thus
finding an optimal code for such penalties is more difficult.  There
should, however, be cases in which this is possible for convex
$\varphi$ which grow more slowly than some exponential.

Another extension of this work would be to find coding algorithms for
other probability mass functions under the nonlinear penalties already
considered, e.g., to attempt to use the techniques of
\cite[pp.~103--105]{Baer} for a more reliable algorithm.  Other
possible extensions and generalizations involve variants of geometric
probability distributions; in addition to the one we mentioned that is
analogous to Proposition~(2) in \cite{Abr1}, there are others in
\cite{MSW, GoMa, BCSV}.  Extending these methods to nonbinary codes
should also be feasible, following the approaches in \cite{Abr1} and
\cite{KHN}.  Finally, as a nonalgorithmic result, it might be
worthwhile to characterize {\it all} optimal codes --- not merely
finding {\it an} optimal code --- as in \cite[p.~289]{Goli2}.

\section*{Acknowledgments}

The author wishes to thank the anonymous reviewers, David
Morgenthaler, and Andrew Brzezinski for their suggestions in improving
the rigor and clarity of this paper.

\appendices
\section{Optimal Maximal Redundancy Golomb Codes for Large~$\theta$}
\label{maxred}

Let us calculate optimal maximal redundancy as a function of $\theta
\geq 0.5$:
$$
\begin{array}{rcl}
R^*(\biglen_\theta^{**},\bigp_\theta) 
&=& \max_i \len_\theta^{**}(i) + \lg \boldp_\theta(i) \\
&=& 1 + 
\left\lceil \lg \lceil - \frac{1}{\lg \theta} 
\rceil \right\rceil + \\
&&\lg (1-\theta) + \\
&&\left(2^{\left\lceil \lg \lceil - \frac{1}{\lg \theta}
\rceil \right\rceil} - \left\lceil - \frac{1}{\lg \theta} \right\rceil
\right)\lg \theta \\
&=& 1 - \left\lceil -\frac{1}{\lg \theta} \right\rceil \lg \theta + \\
&&\lg \left( - \frac{1-\theta}{\lg \theta} \right) - \\
&& 2^{\left\lceil \lg \left(- \frac{1}{\lg \theta}\right)
\right\rceil - \lg (- \frac{1}{\lg \theta})} + \\
&&\left\lceil \lg \left(- \frac{1}{\lg \theta}\right)
\right\rceil - \lg \left(- \frac{1}{\lg \theta}\right) \\
&=& 2 + \lg \left( - \frac{1-\theta}{\lg \theta} \right) - 
\left\lceil -\frac{1}{\lg \theta} \right\rceil \lg \theta - \\
&& 2^{1-\langle \lg (- \frac{1}{\lg \theta})
\rangle} - \left\langle \lg \left(- \frac{1}{\lg \theta}\right)
\right\rangle,
\end{array}
$$
where $\langle x \rangle$ denotes the fractional
part of $x$, i.e., $\langle x \rangle \definedas x - \lfloor x \rfloor$, since
$$\left\lceil \lg \lceil - \frac{1}{\lg \theta} \rceil \right\rceil =
\left\lceil \lg \left( - \frac{1}{\lg \theta} \right) \right\rceil$$
for $\theta > 0.25$ (and thus for $\theta \geq 0.5$).  Using the
Taylor series expansion about $\theta = 1$, we find
$$ 
\lg \left( - \frac{1-\theta}{\lg \theta} \right) = - \lg \lg e -
(\lg \sqrt{e})(1-\theta)+\order((1-\theta)^2)
$$
where $e$ is the base of the natural logarithm.
Additionally,
$$-\left\lceil-\frac{1}{\lg \theta} \right\rceil \lg \theta = 1 +
\order(1-\theta).$$ Note that this actually oscillates between $1$ and
$1+(1-\theta)\lg e$ in the limit, so this first-order asymptotic term
cannot be improved upon.  However, the remaining terms
\begin{equation}
2-2^{1-\langle \lg (- \frac{1}{\lg \theta})
\rangle} - \left\langle \lg \left(- \frac{1}{\lg \theta}\right)\right\rangle 
\label{osc}
\end{equation}
oscillate in the zero-order term.  Assigning $x = \langle \lg (-
1/\lg \theta)\rangle$, we find that (\ref{osc}) achieves
its minimum value, $0$, at $0$ and $1$.  The maximum point is easily
found via a first derivative test.  This point is achieved at $x=1-\lg
\lg e$, at which point (\ref{osc}) achieves the maximum value $1-\lg
e+\lg \lg e$.  
Thus, gathering all terms, 
$$\lim \inf_{\theta \uparrow 1} R^*(\biglen_\theta^{**},\bigp_\theta) = 1-\lg
\lg e = 0.4712336270 \ldots,$$ $$\lim \sup_{\theta \uparrow 1}
R^*(\biglen_\theta^{**},\bigp_\theta) = 2 - \lg e = 0.5573049591 \ldots,$$ 
and, overall,
\begin{eqnarray*}
R^*(\biglen_\theta^{**},\bigp_\theta)
&=& 3 - \lg \lg e - \\
&& 2^{1-\langle \lg (- \frac{1}{\lg \theta})
\rangle} - \left\langle \lg \left(- \frac{1}{\lg \theta}\right)
\right\rangle+ \\
&& \order(1-\theta).
\end{eqnarray*}
This oscillating behavior is similar to that of the average redundancy
of a complete tree, as in \cite{Gall} and \cite[p.~192]{Knu3}.
Contrast this with the periodicity of the minimum {\it average}
redundancy for a Golomb code:\cite{Szpa}
\begin{eqnarray*}
\bar{R}(\biglen_{\theta,1}^*,\bigp_\theta) 
&=& 1 - \lg \lg e - \lg e + \\
&& 
2^{2-2^{1-\langle \lg (-\frac{1}{\lg \theta}) \rangle}} 
- \left\langle \lg \left(-\frac{1}{\lg
  \theta}\right) \right\rangle + \\
&&\order(1-\theta)
\end{eqnarray*}
where $\biglen_{\theta,1}^*$ is the optimal code for the traditional
(linear) penalty.

\section{Glossary of Terms}
\label{glossary}

\tablefirsthead{\hline \multicolumn{1}{c}{Notation}
                     & \multicolumn{1}{l}{~Meaning} \\ \hline }
\tablehead{\hline \multicolumn{2}{l}{\small \sl ~~continued}\\
           \hline \multicolumn{1}{c}{Notation}
                     & \multicolumn{1}{l}{~Meaning} \\ \hline }
\tabletail{}
\tablelasttail{}
\begin{supertabular}{l|l}
$a$ & Base of exponential penalty \\
$b(x,k)$ & $(x+1)$th codeword of complete binary code \\
& with $k$ items (i.e., the order-preserving \\
& [alphabetic] code having the first $2^{\lceil \lg k \rceil}-k$ \\
& items with length $\lfloor \lg k \rfloor$ and the last \\
& $2k - 2^{\lceil \lg k \rceil}$ items with length $\lceil \lg k \rceil$) \\
$c(i)$ & Codeword (for symbol) $i$ \\
$C$ & Code $\{c(i)\}$ \\
$e$ & Base of the natural logarithm ($e \approx 2.71828$) \\
G$\kval$ & Golomb code with parameter $\kval$, one of the \\
&form $\{1^{\lfloor j/\kval \rfloor} 0 b(j \bmod \kval, \kval) : j \geq 0\}$ \\
$H_{\alpha}(\bigp)$ & R\'{e}nyi entropy $(1-\alpha)^{-1} \lg \sum_{i \in \X} 
\boldp(i)^{\alpha}$ \\
& (or, if $\alpha \in \{0,1,\infty\}$, the limit of this) \\
$i^{**}$ & Index of the codeword that, among a \\
& given code's inputs $i \in \X$, maximizes \\
& pointwise redundancy, $\len(i)+\lg \boldp(i)$ \\
$j \bmod k$ & $j-k \lfloor j/k \rfloor$ \\
$\CampCost_a (\bigp,\biglen)$ & Penalty $\log_a \sum_{i\in \X} \boldp(i) a^{\len(i)}$ \\
$\len(i)$ & Length of codeword (for symbol) $i$ \\
$\biglen$ & $\{\len(i)\}$, the lengths for a given code \\
$\len^{(r)}(i)$ & Length of codeword $i$ of an optimal code \\
& minimizing maximum redundancy for $\bigw^{(r)}$ \\ 
$\biglen^{(r)}$ & $\{\len^{(r)}(i)\}$, the lengths of an optimal code \\
& minimizing maximum redundancy for $\bigw^{(r)}$  \\
$\len^*(i)$ & Length of codeword $i$ of an optimal code \\
& for an exponential penalty, $\CampCost$ \\
$~~(\len_{\theta,a}^*(i))$ & ~~(...if $\theta$ and $a$ are specified) \\
$\biglen^*$ & $\{\len^*(i)\}$, the lengths of an optimal code \\
$~~(\biglen_{\theta,a}^*)$ & ~~(...if $\theta$ and $a$ are specified) \\
$\len_{\theta,a,d}^*(i)$ & Length of codeword $i$ of an optimal code \\
& minimizing $d$th exponential redundancy \\
$\biglen_{\theta,a,d}^*$ & $\{\len_{\theta,a,d}^*(i)\}$, the lengths of an optimal code\\
& minimizing $d$th exponential redundancy \\
$\len^{**}(i)$ & Length of codeword $i$ of an optimal code \\
& minimizing maximum redundancy \\
$\biglen^{**}$ & $\{\len^{**}(i)\}$, the lengths of an optimal code \\
& minimizing maximum redundancy \\
$\order(\cdot)$ & Order of $\cdot$ asymptotic complexity \\
$\boldp(i)$ & Probability of input symbol $i$ \\
$~~(\boldp_\theta(i))$ & ~~(...for geometric dist\textsuperscript{r} with parameter $\theta$) \\
$~~(\boldp_\lambda(i))$ & ~~(...for Poisson dist\textsuperscript{r} with parameter~$\lambda$) \\
$\bigp$ & $\{\boldp(i)\}$, the input probability mass function \\
$~~(\bigp_\theta)$ & ~~(...for geometric dist\textsuperscript{r} with parameter $\theta$) \\
$~~(\bigp_\lambda)$ & ~~(...for Poisson dist\textsuperscript{r} with parameter~$\lambda$) \\
$\bar{R}_a(\biglen, \bigp)$&$\CampCost_a (\bigp,\biglen) - H_{\alpha(a)}(\bigp)$, the average \\
&pointwise redundancy \\
$R_d(\biglen, \bigp)$&$d^{-1} \lg \sum_{i \in \X} \boldp(i) 2^{d\left(\len(i)+\lg \boldp(i)\right)}$, \\
&the $d$th exponential redundancy \\
$R^*(\biglen, \bigp)$&$\max_{i \in \X} [\len(i)+\lg \boldp(i)]$, the maximum \\
& pointwise redundancy \\
$\R$ & The set of real numbers \\
$\Rp$ & The set of positive real numbers \\
$s_0$ & Upper bound on $s^*$ \\
$s^*$ & $\ln a$ for $a$ corresponding to optimal coding \\
& for buffer overflow \\
$\boldw(i)$ & Weight (for symbol) $i$ \\
$\bigw$ & $\{\boldw(i)\}$, the set of weights \\
$\boldw^{(r)}(i)$ & $\boldp(i)$ for $i \leq r$, $2\boldp(r+1)$ for $i=r+1$ \\
$\bigw^{(r)}$ & $\{\boldp(0), \boldp(1), \ldots, \boldp(r), 2\boldp(r+1)\}$ \\
$\X$ & Input alphabet (usually $\X_\infty = \{0, 1, \ldots \}$) \\
$\alpha(a)$ & $1/(1+\lg a)$ (parameter for R\'{e}nyi entropy) \\
$\theta$ & Geometric dist\textsuperscript{r} parameter ($\boldp_\theta(i) = (1-\theta)\theta^i$) \\
$\lambda$ & Poisson dist\textsuperscript{r} parameter ($\boldp_\lambda(i)=\lambda^i e^{-\lambda}/i!$) \\
$\Phi$ & Golden ratio, $\frac{1}{2}(1+\sqrt{5})$ \\
\end{supertabular}

\ifx \cyr \undefined \let \cyr = \relax \fi

\end{document}